\documentstyle[11pt]{report}
\def\bra#1{\mathopen{\langle#1\,|}}

\def\ket#1{\mathclose{|\,#1\rangle}}

\newcommand{\qaqs}{
\setlength{\unitlength}{0.5mm}
\raisebox{-4.2mm}{
\begin{picture}(40,20)(-15,-10)
\put(-19,10){\makebox(0,0){$\bar q$}}
\put(-19,-10){\makebox(0,0){$q$}}
\put(19,-10){\makebox(0,0){$q$}}
\put(19,10){\makebox(0,0){$\bar q$}}
\put(-15,-10){\vector(3,2){9}}
\put(-7.5,-5){\line(3,2){18}}
\put(15,10){\vector(-3,-2){9}}
\put(-15,10){\line(3,-2){9}}
\put(0,0){\vector(-3,2){9}}
\put(0,0){\vector(3,-2){9}}
\put(15,-10){\line(-3,2){9}}
\put(0,0){\circle*{4}}
\end{picture}}}

\newcommand{\qqs}{
\setlength{\unitlength}{0.5mm}
\raisebox{-4.2mm}
{\begin{picture}(40,20)(-15,-10)
\put(-19,-10){\makebox(0,0){$q$}}
\put(-19,10) {\makebox(0,0){$q$}}
\put(19,-10) {\makebox(0,0){$\bar q$}}
\put(19,10)  {\makebox(0,0){$\bar q$}}
\put(-15,-10){\vector(3,2){9}}
\put(-7.5,-5){\vector(3,2){18}}
\put(7.5,5)  {\line(3,2){9}}
\put(-15,10) {\vector(3,-2){9}}
\put(-7.5,5) {\vector(3,-2){18}}
\put(7.5,-5){\line(3,-2){9}}
\put(0,0){\circle*{4}}
\end{picture}}}
    
\begin{document}
    \title{\huge\bf Possible Phases of the Strong Interaction
           Vacuum\footnote{Talk given at the {\em Multiplicity
           Production and Heavy Ion Collision} Workshop held at  CCAST
           (World Laboratory), Beijing, Mar. 4-9, 1996}}
    \author{Shuqian Ying\\
            Physics Department, Fudan University\\
            Shanghai 200433, China\\
             and \\
            CCAST (World Laboratory), P. O. Box 8730 \\ Beijing
            100080, China}
    \maketitle
       \begin{abstract}
             A study of the possible vacuum phases in a strongly
             interacting two flavor light quark system is presented.
             Four possible phases are found with some of their
             properties presented. The possible existence of a
             localized diquark condensed phases inside
             a nucleon and excited hadrons is investigated by showing the
             potential of solving three selected current ``puzzles'' in
             experimental obervations if such a possibility is taken into
             account. Base on these results, the potential of
             producing these phases in heavy ion collisions is speculated.
       \end{abstract}
    \tableofcontents
     \chapter{A General Introduction}
    The strong interaction vacuum is believed to undergo certain kind
of phase transition in high temperature and density environments. Such a
belief at high temperature is substantiated by both lattice QCD
computations \cite{LatQCD} and the statistical
bootstrap argument \cite{Bootstrap} of the early days when the fundamental
particles of strong interaction are considered to be baryons, mesons and
corresponding resonances. The destination phases of the transition from
the present day large scale strong interaction vacuum phase, in which
the temperature and density can be regarded as zero,
and the nature of these transitions remain, however, poorly understood. The
best known new phase of QCD is the one called quark--gluon plasma
(QGP), which was found on
lattice QCD simulations \cite{LatQCD}. Other phases, such as the
chiral symmetry restoration phase \cite{Chiral-P}, was also shown to
happen together with the QGP in lattice QCD calculations.

  To understand these questions is a worthwhile endeavor since, on the one
hand, the behaviors of QCD are not known well enough to have all its
possible phases listed with their properties predicted
and, on the other hand, the experimental
probes of the hadronic matter in the laboratories and the analysis of
resulting data have not provided us with sufficiently clear pictures
for the nature of the vacuum
phases produced. For example, it is still not known for certain whether
or not the QGP has been created in heavy ion collisions. However, an
understanding of the above mentioned problems is of fundamental
importance since the knowledge gained in this area can help us to
understand the evolution of the early universe, the properties of
the stellar matters in
heavy stars, and eventually to tell, e.g., where matter come from,
why there appears to be an arrow for the time, etc..
It thus constitutes one of the frontiers of our knowledge that awaits
to be explored.

  QCD is a highly nonlinear theory that is difficult to solve,
especially at low and intermediate energies. Lattice simulations are
still limited by the power of computers, which allows the computation only on
small lattice (size) systems. In addition, fermions are not easy to handle on
lattices. The approach adopted by this study for the theoretical
explorations is to use simplified
fermionic models for the strong interaction which inherit most of the
basic symmetries of QCD at long distances. These models can be
considered to have large overlap with the ones that are derivable by
integrating the gluon degrees of freedom (in the path integration
sense) out from the original QCD Lagrangian density. Despite such an
expectation may in fact not easily fulfilled in reality, we still
consider this
undertaken valuable since it can provide sharp predictions that can be
compared to our empirical knowledge due to the relative simplicity of
the model approaches. In addition, the simplicity of the model
approach can let us investigate large set of possibilities (for the
fermionic sector) that is otherwise not possible. In contrast to the
theoretical exploration efforts, the model dependency of the study
related to realistic physical systems, in which the theoretical
possibilities is searched for, is reduced as much as possible by using
model independent methods like, e.g., symmetry constraints, sum rules,
etc.. In doing so, the conclusion derived can be more reliable.

On the observational sides, the information about a nucleon in various
reactions, the multiplicity production in $e^+ e^-$ annihilation and
many others can be used to determine whether the theoretical
possibilities discovered in model studies actually exist inside static
systems like a nucleon and in a time dependent system like in the
$e^+e^-$ annihilation system. Albeit a negative result of the above
mentioned investigations does not necessary mean that these possibilities are
forbidden in other processes like
heavy ion collisions, which are the main topic of this workshop, the early
universe, etc., a positive result from these ``domestic'' experiments
does increase the probabilities and are thus worth of studying.
In addition, the structure of a nucleon, the hadronization mechanism
in the $e^+ e^-$ annihilation are interesting problems on their own
rights.

The report consists of two parts. In the first part, which is based
on Refs. \cite{YingL,YingA} and work undertaken Ref.
\cite{YingB}, two massless fermionic models for the strong interaction
are introduced with their vacuum phase structure determined. The
massless fermions are identified as up (u) and down (d) quarks. Four
kinds of phases are found: 1) the bare vacuum 2) the so called
$\alpha$-phase, in which the original chiral $SU(2)_L\times
SU(2)_R$ symmetry is spontaneously broken down to a flavor $SU(2)_V$
symmetry, that is widely studied in literature 3) the $\beta$-phase,
in which the original chiral $SU(2)_L\times
SU(2)_R$ symmetry is also spontaneously broken down to a flavor $SU(2)_V$
symmetry; it spontaneously breaks the $U(1)$ symmetry corresponding
to baryon number conservation induced by a condensation of diquarks 4)
the $\omega$-phase, in which the original chiral $SU(2)_L\times
SU(2)_R$ symmetry is unbroken. The $U(1)$ symmetry corresponding
to baryon number conservation is also spontaneously broken,
like in the $\beta$-phase.
The second part consists of a study of three different issues. The
first one is related to an examination of the PCAC relationship for a
nucleon, which is based upon Ref. \cite{Epcac}. Certain
inconsistency that is in favor of the existence of a localized $\beta$-phase
inside a nucleon is revealed. The
second one provides a gauge invariant modification of the
Gerasimov-Drell-Hearn (GDH) sum rule required by current experimental data. It
is based upon Ref. \cite{GDH}. If the deviation of the experimental
data from the GDH sum rule is proven genuine, a spontaneous broken
down of the $U(1)$ symmetry correspond to baryon number conservation
inside a nucleon is shown to be a necessity. One of the most natural
realization of such a symmetry breaking is by having $\beta$- or
$\omega$- phase inside a nucleon. The third one is based upon the observed
baryon--antibaryon rapidity difference correlation in a high energy $e^+e^-$
annihilation which is discussed by Prof. Xie's group \cite{Xie} in Shandong
University in this workshop.
Experimental evidence exists \cite{OPAL} that baryon number and
antibaryon number is not locally produced in the observed jets. Such a
violation of locality can be shown to be a direct consequence
of the fact that the string that fragments in to hadrons contains
either the $\beta$- or $\omega$- phase. In the last chapter of this
report, a summary is provided and
some speculations concerning the possibility of producing the
$\beta$- or/and $\omega$- phases is given.
\part{ Theoretical Exploration}
\chapter{Preliminaries}
\section{Introduction}
The basic participants of the strong interaction we know of today are
color carrying quarks (antiquarks) and gluons with basic interaction
between them mediated by gluons according to the
QCD Lagrangian density, which is a non-Abelian gauge theory with
asymptotic freedom at short distances or high energies. At low and
intermediate energies, the interaction between color sources
(including quarks and gluons) becomes strong enough to render a
perturbative analysis useless. For a light quark system, strong
interaction can cause quark--antiquark pair production from the bare
vacuum. With the increase of the interaction strength, it is expected
that a macroscopic number of quark--antiquark pairs
can be produced resulting in effects that survive the thermal limit
to lead to various phase transitions in the vacuum.

   At a formal level, the gluon degrees of freedom can be integrated
out to obtain a pure quark--quark interaction effective action.
Such an effective quark--quark interaction action are expected
to be very complicated if can be practically computed at all and
difficult to analyze to extracting useful information without certain
physical picture being built up using other indirect approaches. One
of these approaches is to build models for the quark--quark interaction
part that inherits the basic symmetries of the original Lagrangian
density. It is expected that the major physics emerged from these models
has large overlap with the one implied by the QCD Lagrangian.

  Before list the symmetries considered, let's make the first simplification.
The behaviors of certain massless fermion system are expected to
represent the light quark system, which is restricted to the up (u) and
down (d) quark sector in the flavor space here, well.
Thus, for the simplicity of the discussion, we shall assume that the
mass of the u and d quarks is zero. QCD Lagrangian density with massless u and
d
quarks has, besides the ones listed in the following, a $SU(2)_L\times
SU(2)_R$ chiral symmetry. So the symmetries that these models should
have in addition are
1) invariance under the Lorentz group 2) global $U(1)$ symmetry
corresponding to the conservation of baryon number 3) local $U(1)_{em}$
symmetry
corresponding to the electromagnetism 4) global color $SU(3)_c$ symmetry.

One of the well studied model with the above properties is the Nambu
Jona--Lasinio (NJL) model \cite{NJLm} used at the light quark level
(some include the strange quark as well).

\section{Auxiliary field method using a simple model as an example}

   The fermion--fermion interaction terms are assumed to be of
contact 4--fermion form in the first part of the
report. One of the best methods of treating these nonlinear 4--fermion
interaction models is to introduce auxiliary fields
\cite{AuxF,GNmodel}. The subtleties
of introducing auxiliary fields related to the Fierzing ${\cal F}$
and Crossing ${\cal C}$ operations are discussed in more detail in
Ref. \cite{YingA}. It shall not be discussed here. Suffice
it will be to use the following heuristic steps to show how
auxiliary fields are introduced to handle these non-linear terms.

   For simplicity, let's consider the following fermion--fermion
interaction model
\begin{eqnarray}
    {\cal L} &=& \bar\psi \left ( i\rlap\slash\partial - m \right )
    \psi + G_0 \left (\bar\psi\psi \right )^2,
\label{SimpMod}
\end{eqnarray}
where $m$ is the mass of the fermion,
$\psi$ is a 4--component Dirac spinor fermion field and $G_0$ is
the coupling constant.

 In the path integration formalism, the generating functional
$W[\eta,\bar\eta]$ for the connected Green functions for the fermions is
\begin{eqnarray}
      \exp\left (i W[\eta,\bar\eta] \right ) &=& N
        \int D[\psi] D[\bar\psi] \exp \left [i
               \int d^4x \left ( {\cal L} + \bar \eta \psi + \bar\psi\eta
            \right ) \right ],
\label{Gen-Func1}
\end{eqnarray}
where $\eta$, $\bar\eta$ are external Grassmanian fields,
$N$ is a constant that are of no physical effects; it is so chosen
that $W[0,0]$ is zero.

Since the fermion fields are not bilinear in the
arguement of the exponential integrand, the functional integration
over the fermion fields can not be easily computed. A step forward
can be achieved if the following mathematical transformation is used,
namely,
\begin{eqnarray}
      \exp\left (i W[\eta,\bar\eta] \right ) &=& N'
        \int D[\psi] D[\bar\psi] \exp \left [
        i\int d^4x \left ( {\cal L} + \bar \eta \psi + \bar\psi\eta
             \right ) \right ] \nonumber \\
            &&\times \int D[\sigma] \exp \left [-i\int d^4x
             (a \sigma + b\bar\psi\psi)^2
             \right ],
\label{Gen-Func2}
\end{eqnarray}
where $\sigma$ is the auxiliary field introduced and $a$, $b$ are
arbitrary constant to be determined in the following. The second
functional integration over $\sigma$ gives an $\eta$ and $\bar\eta$
independent constant contribution since it is a complete square in the
arguement of the exponential been integrated; it causes no physical
consequences since its effects can be absorbed in to the normalization
constant $N'$. After writing the generating functional in this form,
progress can be made by a proper choice of the constants $a$ and
$b$. Let $a= 1/2\sqrt{G_0}$ and $b=\sqrt{G_0}$, then Eq. \ref{Gen-Func2} is
\begin{eqnarray}
      \exp\left (i W[\eta,\bar\eta] \right ) &=& N'
        \int D[\sigma] \int D[\psi] D[\bar\psi] \exp \left [i
        \int d^4x \left ( {\cal L}' + \bar \eta \psi + \bar\psi\eta
             \right ) \right ]
\label{Gen-Func3}
\end{eqnarray}
with
\begin{eqnarray}
    {\cal L}' &=& \bar\psi \left (i\rlap\slash\partial - \sigma - m
  \right )\psi - {1\over 4 G_0} \sigma^2.
\label{Eff-L1}
\end{eqnarray}
When the space--time dimension is taken to be $1+1$, this is the
half bosonized Gross--Neveu model \cite{GNmodel}.
It can be seen that the 4--fermion term in the original
Lagrangian density are absent in the new one; this
allows the integration over the fermion fields be carried out. The
price that one should pay is to introduce additional field,, namely the
auxiliary field $\sigma$, to be functionally integrated.
Eq. \ref{Gen-Func3} becomes
\begin{eqnarray}
      \exp\left (i W[\eta,\bar\eta] \right ) &=& N'
        \int D[\sigma] \exp \left ( i S_{eff}[\sigma]+
        \bar\eta S_F \eta  \right ),
\label{Gen-Func4}
\end{eqnarray}
where the effective action $S_{eff}$ for the auxiliary field $\sigma$
after integrate the fermion fields can be expressed as
\begin{eqnarray}
     S_{eff}[\sigma] &=& -i \mbox{Sp} \mbox{Ln}  \gamma^0 \left
                 (i\rlap\slash\partial-\sigma - m \right )
                 - { 1\over 4 G_0} \int d^4x \sigma^2,
\label{EffAct1}
\end{eqnarray}
with ``Sp'' denoting the functional trace of the corresponding
operator. Formally, $S_{eff}$ can be expressed in terms of the
eigenvalues $\lambda$ of operator $\hat D\equiv \gamma^0
 (i\rlap\slash\partial -\sigma - m)$, namely,
\begin{eqnarray}
    S_{eff}[\sigma] &=& -i\sum_\lambda \ln {
         \lambda[\sigma]\over \lambda[0] }
          - {1\over 4 G_0} \int d^4x\sigma^2,
\label{EffAct2}
\end{eqnarray}
where $\lambda[\sigma]$ is the eigenvalue of $\hat D$ and the
summation is over all of the eigenvalues considered.

In case of $\sigma$ independent of space--time, which is the case for
the vacuum state of the system, Eq. \ref{EffAct2} can
be simplified further to
\begin{eqnarray}
    S_{eff}(\sigma) &=& -i V_4 \int {d^4p\over (2\pi)^4} \ln
          \left [ { p^2 - (\sigma + m)^2 \over p^2- m^2} \right ]
         - {1\over 4 G_0} V_4 \sigma^2,
\label{EffAct3}
\end{eqnarray}
where $V_4\to \infty$ is the space--time volume of the system. The
vacuum $\sigma$ value of the system can be determined by minimizing
the effective potential $V_{eff}(\sigma) = -S_{eff}(\sigma)/V_4$.

\section{8-component Dirac spinor}

 The best representation for the Dirac spinor for the path integral
formalism developed in this
study is the 8-component ``real'' one \cite{YingL,YingA,YingB}
as oppose to the 4-component
one. An 8-component ``real'' Dirac spinor can be written as
\begin{eqnarray}
      \Psi & = & \left ( \begin{array}{c} \psi_1 \\ \psi_2 \end{array}
    \right ) ,\label{8DimRep1}
\end{eqnarray}
where $\psi_1$ and $\psi_2$ are 4-component Dirac spinors. They are
related to each other in the following way
\begin{eqnarray}
     \psi_2 & = & \left \{  \begin{array} {lr}
                                C \bar\psi_1^T &\mbox{\hspace{0,8in}One
flavor}\\
                                Ci\tau_2 \bar \psi_1^T &
                                \mbox{\hspace{0.8in} Two
                                  flavor}
                            \end{array}
        \right .
\label{PsiRealRel}
\end{eqnarray}
where $C=-C^{-1}$ is the charge conjugation operator (in the
4-component representation of the Dirac spinor) and $\tau_2$ is the
second Pauli matrices acting on the flavor space of the fermion field
$\Psi$.

For later discussions, it proves useful to introduce three Pauli
matrices $O_{1,2,3}$ acting on the two 4-component Dirac spinor $\psi_{1,2}$.
With $O_i$, Eq. \ref{PsiRealRel} can be more compactly written as
\begin{eqnarray}
   \bar\Psi &=&\Psi^T \Omega\label{Psi},\label{PsiBar}
\end{eqnarray}
with the $\Omega$ matrix defined as
\begin{eqnarray}
\Omega&=& \left (\begin{array}{cc} 0 & - C^{-
1} \rho^{-1}\\ C \rho &
                  0\end{array}\right ),
\label{G}
\end{eqnarray}
where $\rho = \rho^{-1} = 1$ for one flavor case and $\rho = i\tau_2 =
- \rho^{-1}$ for two flavor case.

Using the 8-component $\Psi$, the Lagrangian density corresponding to
Eq. \ref{Eff-L1} takes to following form
\begin{eqnarray}
     {\cal L}' &=& {1\over 2} \bar\Psi \left (
          i\rlap\slash\partial - \sigma - m \right ) \Psi
         - {1\over 4 G_0} \sigma^2.
\label{Eff-L1-8}
\end{eqnarray}
The effective action for $\sigma$ is then
\begin{eqnarray}
     S_{eff}[\sigma] &=& -{i\over 2} \mbox{Sp} \mbox{Ln}  \gamma^0 \left
                 (i\rlap\slash\partial-\sigma - m \right )
                 - { 1\over 4 G_0} \int d^4x \sigma^2+const
\label{EffAct1-8}
\end{eqnarray}
due to Eq. \ref{PsiBar}. Here ``$const$'' is chosen such that
$S_{eff}[0]=0$.

\chapter{Models}
\section{Classification of the 4--fermion interaction terms}

  The light quark system in reality contains fermionic u and d quarks with
three
colors. Therefore the fermion fields that should be used (in the
4-component form) is $4\times 2\times 3=24$ component. For a modeling
of the quark system, a Dirac spinor $\psi_{fc}$ with $f=u,d$ labeling
the flavor and $c=1,2,3$ labeling the color has to be used. For the
compactness, the flavor and color indices of $\psi$ will be suppressed
in the following as long as no confusion is thought to occur.

If the mass of the light quarks is  assumed to be zero, the full
Lagrangian density of the system is written as
\begin{eqnarray}
  {\cal L} &=&  \bar\psi  i\rlap\slash\partial \psi + {\cal L}_{int}.
\label{Lagrangian}
\end{eqnarray}
The 4--fermion interaction terms can be generally classified into two
categories in the quark--antiquark channel
\begin{eqnarray}
{{\cal L}}_{int} & = & \left. \qaqs \right\}\begin{array}{c}
color\\singlet\end{array}+\left. \qaqs \right\}
\begin{array}{c}
     color\\octet
\end{array} +
Fierz\hskip 0.08in
		 term\nonumber\\
	   \nonumber\\
	   & = & {{\cal L}}^{(0)}_{int} +
{{\cal L}}^{(8)}_{int},\label{LAG1}
\end{eqnarray}
where ${{\cal L}}^{(0)}_{int}$ generates quark--antiquark
scattering in color
singlet
channel and ${{\cal L}}^{(8)}_{int}$ generates quark--antiquark
scattering in color octet channel.

For ${{\cal L}}^{(0)}_{int}$, the well
known two flavor chiral symmetric Nambu Jona--Lasinio (NJL) \cite{NJLm}
interaction is
chosen, namely
\begin{eqnarray}
{{\cal L}}^{(0)}_{int} & = & G_0 \left [ (\bar\psi\psi)^2 +
(\bar\psi i\gamma^5\vec{\tau}\psi)^2 \right ].
\label{NJLL}
\end{eqnarray}

The color octet ${\cal L}^{(8)}$ is written in the quark--quark
(antiquark--antiquark) channel form for our purposes, namely
\begin{eqnarray}
{{\cal L}}^{(8)}_{int} & = & \left. \qqs \right
\}\begin{array}{c}
				 color \\
				 triplet \end{array} +
\left. \qqs\right\}
				 \begin{array}{c}
				 color \\ sextet
				 \end{array}
		  + (q\leftrightarrow \bar q)\nonumber\\
		  & = &{{\cal L}}^{(3)}_{int} +
{{\cal L}}^{(6)}_{int}.\label{lag8}
\end{eqnarray}
The color sextet term is repulsive in the one gluon exchange case.
Due to the non-existence of colored baryon containing three quarks
in nature, it is assumed to be generally true. So we restrict
ourselves to the
attractive color triplet two quark
interaction terms. The attractive color triplet quark bilinear terms
can be classified according to their transformation
properties under Lorentz and
chiral
$SU(2)_L\times SU(2)_R$ groups. In general,
if only terms without derivative in fermion fields are
considered, ${\cal L}_{int}^{(3)}$ has the following
form
\begin{eqnarray}
{{\cal L}}^{(3)}_{int} &=&{1\over 2}\sum_r G_r\sum_{ab} C^r_{ab}
(\bar\psi
       \Gamma_a^r\tilde{\bar\psi})(\tilde\psi\Gamma^r_b\psi),\label{
L3}
\end{eqnarray}
with $\Gamma^r_a$, $\Gamma^r_b$ matrices in Dirac, flavor and color
spaces generating representation ``$r$'' and satisfying the
antisymmetrization condition
\begin{eqnarray}
\left (\Gamma^r_{a,b}i\tau_2C\right )^T & = &
-\Gamma^r_{a,b}i\tau_2 C,\label{Gamma}
\end{eqnarray}
for the fermionic systems.

 Operator $\tilde\psi\Gamma^r_b\psi$ belongs to
an irreducible representation ``r'' of chiral, Lorentz and color groups and
$\bar\psi\Gamma^r_a\tilde{\bar\psi}$ belongs to the conjugate
representation.
Coefficients $C_{ab}^r$ render the summation $\sum_{ab}\ldots$ invariant
under Lorentz, chiral
$SU(2)_L\times SU(2)_R$ and color $SU(3)_c$ groups. $\{G_r\}$ is a
set of independent
4--fermion coupling constants characterizing the color
triplet quark--quark
interactions. The tilded fermion field operators $\tilde\psi$
and $\tilde{\bar\psi}$ are defined as
\begin{eqnarray}
\tilde\psi & = & \psi^T(-i\tau_2)C^{-1},\\
\tilde{\bar\psi} & = & Ci\tau_2\bar\psi^T.
\end{eqnarray}

 The transformation properties of a list of bilinear products of
two fermion fields in color triplet and sextet representations are given
in Table 1, where all of the possible ones without any derivative in fermion
fields are included.
\begin{table}[h]
\caption{
Transformation properties of various bilinear fermion operators under
the action of
Lorentz, chiral $SU(2)_L\times SU(2)_R$ and color $SU(3)_c$ groups.
\label{Table1}}
\begin{center}
\begin{tabular}{|c|ccccc|}
\hline
     Representation & Operators & Lorentz\ Group &$SU(2)_L\times SU(2)_R$&
						    $SU(3)_c$&\phantom{abc}\\
\hline
&&&&&\\
 1   & $\tilde\psi\psi$ & $Pseudoscalar$ & 1 $dim$ & $\bar 3$ &\\
&&&&&\\
 2   & $\tilde\psi\gamma^5\psi$ & $Scalar$ & 1 $dim$ & $\bar 3$ &\\
&&&&&\\
 3 & $\left ( \matrix{\tilde\psi{\bf\tau}\psi\\
    \tilde\psi\gamma^5{\bf\tau}\psi\\}\right ) $
  & $\matrix{Pseudoscalar\\ Scalar\\}$ & 6\ $dim$ &
  $\matrix{\bar 6\\ \bar 6\\}$ &\\
&&&&&\\
 4 & $\left ( \matrix{\tilde\psi\gamma^\mu\psi\\ \tilde\psi\gamma^\mu\gamma^5
       {\bf\tau}\psi\\}\right ) $ & $\matrix{Axial\ Vector\\ Vector\\}$ & 4\
$dim$
	     &    $\matrix{\bar 6\\ \bar 6\\}$ &\\
&&&&&\\
 5 & $\left ( \matrix{\tilde\psi\gamma^\mu\gamma^5\psi\\ \tilde\psi\gamma^\mu
       {\bf\tau}\psi\\}\right )$ & $\matrix{Vector\\ Axial\ Vector\\}$ & 4\
$dim$
	     &     $\matrix{\bar 3\\ \bar 3\\}$ &\\
&&&&&\\
 6 & $\tilde\psi\sigma^{\mu\nu}\psi$& $Tensor-$ & 1\ $dim$ & $\bar 6$ &\\
&&&&&\\
 7 & $\left ( \matrix{\tilde\psi\sigma^{\mu\nu}{\bf\tau}\psi\\ \tilde\psi
       \sigma^{\mu\nu}\gamma^5
	     {\bf\tau}\psi\\}\right )  $ & $\matrix{Tensor-\\ Tensor+\\}$ & 6\
$dim$
	     & $\matrix{\bar 3\\ \bar 3\\}$ & \\
&&&&&\\
\hline
\end{tabular}
\end{center}
\end{table}

\section{Model 1 and its vacuum phase diagram}

   The 4--fermion interaction terms are classified in the above
section. The representation
``2'' of table \ref{Table1} is chosen to form the first model
interaction term. In terms of the 8-component Dirac spinor
$\Psi$ and after half bosonization \cite{YingA,YingB}, it takes the
following form
\begin{eqnarray}
 {\cal L}_1 & = & {1\over 2} \bar \Psi\left [i{\rlap\slash\partial}-\sigma-
             i\vec{\pi}\cdot \vec{\tau}\gamma^5 O_3-\gamma^5 {\cal
             A}_c\chi^c O_{(+)}-\gamma^5 {\cal A}^c\bar\chi_c O_{(-)}
             \right ]\Psi \nonumber \\
            &   &  -{1\over 4 G_0} (\sigma^2 +
             \vec{\pi}^2) + {1\over 2 G_{3'}} \bar\chi_c \chi^c,
\label{Model-L-1}
\end{eqnarray}
where $\sigma$, $\vec{\pi}$, $\bar\chi_c$ and $\chi^c$ are auxiliary
fields, $(\chi^c)^{\dagger} = - \bar\chi_c$,
$G_0$ and $G_{3'}$ are coupling constants of the model.

If the auxiliary fields do not depend on the space-time, the effective
potential can be computed in the momentum space, it has the following
form
\begin{eqnarray}
    V_{eff} &=& 4i\int {d^4p\over (2\pi)^4} \ln\left [\left (
            1- {\sigma^2+\chi^2\over p^2} \right )^2-{\sigma^2\over
              p^2}
              \left ( 1-{\sigma^2-\chi^2\over p^2} \right )^2\right ]
              \nonumber \\
            & & + {1\over 4 G_0}\sigma^2 + {1\over 2G_{3'}} \chi^2,
\label{Veff-2-2}
\end{eqnarray}
where $\chi^2 \equiv -\bar\chi_c\chi^c$.

\begin{figure}[h]
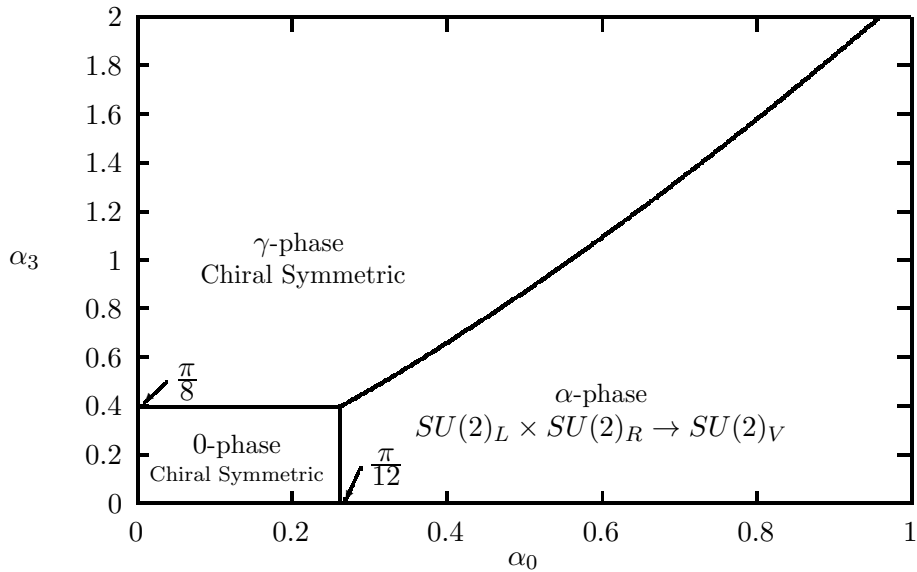

\begin{center}
\setlength{\unitlength}{0.240900pt}
\ifx\plotpoint\undefined\newsavebox{\plotpoint}\fi
\sbox{\plotpoint}{\rule[-0.500pt]{1.000pt}{1.000pt}}%

\end{center}
\caption{\label{Fig:bound1} The phase boundaries between the
  $\alpha$-phase, $\gamma$-phase and the $0$-phase. The chiral
  symmetry is unbroken in both the $\gamma$-phase and the
  $0$-phase. The $\alpha$-phase breaks the chiral symmetry
  spontaneously down to a flavor symmetry.}
\end{figure}

A numerical evaluation in Euclidean momentum space shows that the minima of
$V_{eff}(\sigma,\chi)$ is located on either the $\sigma$ axis
($\chi=0$) or the $\chi$ axis ($\sigma=0$). $V_{eff}(\sigma,0)$ and
$V_{eff}(0,\chi)$ are found to be
\begin{eqnarray}
     v_{eff}(\sigma,0) &=& 3 f({\sigma^2\over\Lambda^2}) + {1\over
       16\pi\alpha_0} {\sigma^2\over\Lambda^2},\label{Veff10}\\
     v_{eff}(0,\chi)   &=& 2 f({\chi^2\over\Lambda^2}) + {1\over
       16\pi\alpha_{3'}} {\chi^2\over\Lambda^2},\label{Veff01}
\end{eqnarray}
where $\Lambda$ is the Euclidean $p^\mu$ space cutoff that defines the
model, the dimensionless effective potential is defined by
$V_{eff}\equiv \Lambda^4 v_{eff}$,
$\alpha_{0} = G_0\Lambda^2/4\pi$ and $\alpha_{3'} =
G_{3'}\Lambda^2/8\pi$ with
\begin{eqnarray}
     f(x) &=& {1\over 8\pi^2}\left [-x + \ln\left (1+{1\over x}\right)
     x^2 - \ln(1+x) \right ]. \label{f-func}
\end{eqnarray}
The value of $V_{eff}$ at the minima of Eqs. \ref{Veff10}
and \ref{Veff01} determine the vacuum of the system in the one loop
Hartree--Fock approximation for the fermions. It
is easy to represent the phase structure of the model by showing it in
the $\alpha_0$--$\alpha_{3'}$ plane, which is given by Fig.
\ref{Fig:bound1}.
Three kinds of phases for the vacuum
are possible. The first phase, which is called the $0$-phase, is the bare
vacuum.
The second phase, or the $\alpha$-phase, has
non-vanishing average value of $\bar\Psi\Psi$; the chiral
$SU(2)_L\times SU(2)_R$ symmetry is spontaneously broken down to a
$SU(2)_V$ flavor symmetric one in this phase. The
third phase, defined as the $\omega$-phase, has non-vanishing diquark
and anti-diquark condensation characterized by a non-vanishing
$\chi^2$; chiral symmetry is unbroken in this phase.

The phase transition across the boundary between the
$0$- and the $\alpha$- phases ($\alpha^c_0=\pi/12$) and the one
between the $0$- and the
$\omega$- phases ($\alpha^c_{3'}=\pi/8$) are of second order.
The phase transition between the
$\alpha$- and the $\omega$- phases ($\alpha_0>\pi/12$ and
$\alpha_{3'}>\pi/8$) is of first order. The Meissner
effects for the electromagnetic field
 are present in the $\omega$-phase. This is discussed in more
detail in Ref. \cite{YingA} for model II in the following. We shall
relegate such a discussion for this model to other work.

\section{Model 2 and its vacuum phase diagram}

The second model interaction Lagrangian density chosen are constructed
from the ones in
representation ``4''
of table \ref{Table1}. In terms of the 8-component Dirac spinor
$\Psi$ and after half bosonization \cite{YingA,YingB}, its Lagrangian
density is
\begin{eqnarray}
{{\cal L}}_2&=& {1\over 2}\bar\Psi \left [ i{\rlap\slash\partial} -
\sigma -
 i\vec{\pi}\cdot\vec{\tau}\gamma^5 O_3 +
O_{(+)} \left (\phi^c_\mu\gamma^\mu\gamma^5
{\cal A}_c
  +\vec{\delta}_\mu^c\cdot\vec{\tau}\gamma^\mu{\cal
A}_c \right ) \right . \nonumber \\ & & \left .- O_{(-)}\left (\bar\phi_{\mu
c}\gamma^\mu\gamma^5{\cal A}^c
  +\vec{\bar\delta}_{\mu c}\cdot\vec{\tau}\gamma^\mu{\cal
A}^c \right) \right ]\Psi - {1\over 4 G_0}(\sigma^2
+ \vec{\pi}^2) \nonumber \\ && - {1\over
2 G_{3'}}
(\bar\phi_{\mu c}\phi^{\mu c} + \vec{\bar\delta}_{\mu
c}\cdot
\vec{\delta}^{\mu c}),\label{Model-L-2}
\end{eqnarray}
where $\bar\phi_{\mu c}$, $\phi_\mu^c$ with $(\phi_\mu^\dagger)_c =
- \bar\phi_{\mu c}$, $\vec{\bar\delta}_{\mu c}$, $\vec{\delta}_\mu^c$
with $(\vec{\delta}_\mu^\dagger)_c = - \vec{\bar\delta}_{\mu c}$ are
auxiliary fields introduced.

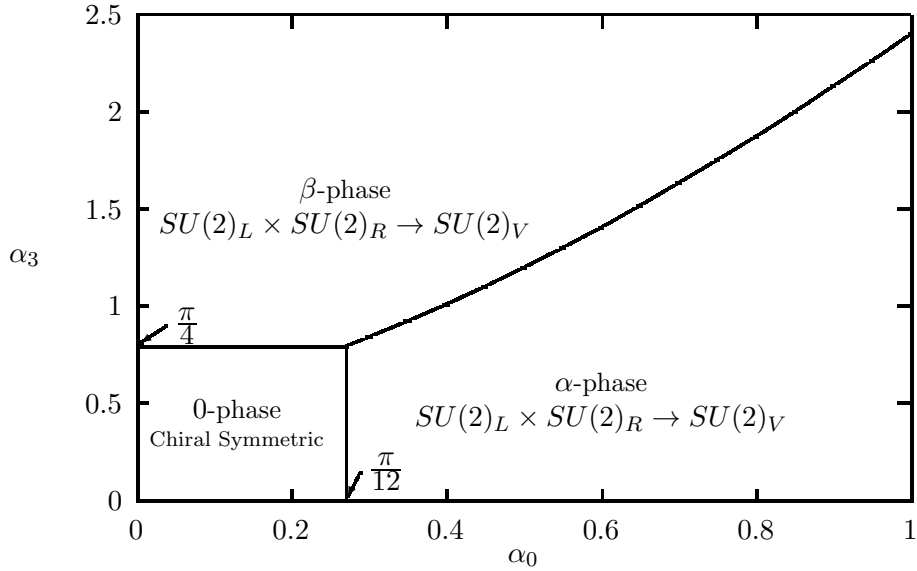
\begin{figure}[h]
\begin{center}
\setlength{\unitlength}{0.240900pt}
\ifx\plotpoint\undefined\newsavebox{\plotpoint}\fi
\sbox{\plotpoint}{\rule[-0.500pt]{1.000pt}{1.000pt}}%
\begin{picture}(1500,900)(0,0)
\font\gnuplot=cmr10 at 10pt
\gnuplot
\sbox{\plotpoint}{\rule[-0.500pt]{1.000pt}{1.000pt}}%
\put(220.0,113.0){\rule[-0.500pt]{292.934pt}{1.000pt}}
\put(220.0,113.0){\rule[-0.500pt]{1.000pt}{184.048pt}}
\put(220.0,113.0){\rule[-0.500pt]{4.818pt}{1.000pt}}
\put(198,113){\makebox(0,0)[r]{$0$}}
\put(1416.0,113.0){\rule[-0.500pt]{4.818pt}{1.000pt}}
\put(220.0,266.0){\rule[-0.500pt]{4.818pt}{1.000pt}}
\put(198,266){\makebox(0,0)[r]{$0.5$}}
\put(1416.0,266.0){\rule[-0.500pt]{4.818pt}{1.000pt}}
\put(220.0,419.0){\rule[-0.500pt]{4.818pt}{1.000pt}}
\put(198,419){\makebox(0,0)[r]{$1$}}
\put(1416.0,419.0){\rule[-0.500pt]{4.818pt}{1.000pt}}
\put(220.0,571.0){\rule[-0.500pt]{4.818pt}{1.000pt}}
\put(198,571){\makebox(0,0)[r]{$1.5$}}
\put(1416.0,571.0){\rule[-0.500pt]{4.818pt}{1.000pt}}
\put(220.0,724.0){\rule[-0.500pt]{4.818pt}{1.000pt}}
\put(198,724){\makebox(0,0)[r]{$2$}}
\put(1416.0,724.0){\rule[-0.500pt]{4.818pt}{1.000pt}}
\put(220.0,877.0){\rule[-0.500pt]{4.818pt}{1.000pt}}
\put(198,877){\makebox(0,0)[r]{$2.5$}}
\put(1416.0,877.0){\rule[-0.500pt]{4.818pt}{1.000pt}}
\put(220.0,113.0){\rule[-0.500pt]{1.000pt}{4.818pt}}
\put(220,68){\makebox(0,0){$0$}}
\put(220.0,857.0){\rule[-0.500pt]{1.000pt}{4.818pt}}
\put(463.0,113.0){\rule[-0.500pt]{1.000pt}{4.818pt}}
\put(463,68){\makebox(0,0){$0.2$}}
\put(463.0,857.0){\rule[-0.500pt]{1.000pt}{4.818pt}}
\put(706.0,113.0){\rule[-0.500pt]{1.000pt}{4.818pt}}
\put(706,68){\makebox(0,0){$0.4$}}
\put(706.0,857.0){\rule[-0.500pt]{1.000pt}{4.818pt}}
\put(950.0,113.0){\rule[-0.500pt]{1.000pt}{4.818pt}}
\put(950,68){\makebox(0,0){$0.6$}}
\put(950.0,857.0){\rule[-0.500pt]{1.000pt}{4.818pt}}
\put(1193.0,113.0){\rule[-0.500pt]{1.000pt}{4.818pt}}
\put(1193,68){\makebox(0,0){$0.8$}}
\put(1193.0,857.0){\rule[-0.500pt]{1.000pt}{4.818pt}}
\put(1436.0,113.0){\rule[-0.500pt]{1.000pt}{4.818pt}}
\put(1436,68){\makebox(0,0){$1$}}
\put(1436.0,857.0){\rule[-0.500pt]{1.000pt}{4.818pt}}
\put(220.0,113.0){\rule[-0.500pt]{292.934pt}{1.000pt}}
\put(1436.0,113.0){\rule[-0.500pt]{1.000pt}{184.048pt}}
\put(220.0,877.0){\rule[-0.500pt]{292.934pt}{1.000pt}}
\put(45,495){\makebox(0,0){$\alpha_3$}}
\put(828,23){\makebox(0,0){$\alpha_0$}}
\put(548,571){\makebox(0,0){\shortstack{$\beta$-phase \\ $\scriptsize
SU(2)_L\times SU(2)_R\to SU(2)_V$}}}
\put(950,266){\makebox(0,0){\shortstack{$\alpha$-phase \\ $\scriptsize
SU(2)_L\times SU(2)_R\to SU(2)_V$}}}
\put(378,232){\makebox(0,0){\shortstack{$0$-phase \\ \scriptsize Chiral
Symmetric}}}
\put(585,159){\makebox(0,0)[l]{$\displaystyle \pi\over \displaystyle 12$}}
\put(281,388){\makebox(0,0)[l]{$\displaystyle \pi\over \displaystyle 4$}}
\put(220.0,113.0){\rule[-0.500pt]{1.000pt}{184.048pt}}
\multiput(570.68,150.06)(-0.496,-0.941){34}{\rule{0.119pt}{2.155pt}}
\multiput(570.92,154.53)(-21.000,-35.528){2}{\rule{1.000pt}{1.077pt}}
\put(552,119){\vector(-1,-2){0}}
\multiput(261.50,385.68)(-0.767,-0.497){46}{\rule{1.806pt}{0.120pt}}
\multiput(265.25,385.92)(-38.252,-27.000){2}{\rule{0.903pt}{1.000pt}}
\put(227,361){\vector(-3,-2){0}}
\multiput(1428.93,843.68)(-0.719,-0.498){76}{\rule{1.702pt}{0.120pt}}
\multiput(1432.47,843.92)(-57.467,-42.000){2}{\rule{0.851pt}{1.000pt}}
\multiput(1367.63,801.68)(-0.755,-0.498){72}{\rule{1.775pt}{0.120pt}}
\multiput(1371.32,801.92)(-57.316,-40.000){2}{\rule{0.888pt}{1.000pt}}
\multiput(1306.74,761.68)(-0.743,-0.498){72}{\rule{1.750pt}{0.120pt}}
\multiput(1310.37,761.92)(-56.368,-40.000){2}{\rule{0.875pt}{1.000pt}}
\multiput(1246.47,721.68)(-0.775,-0.498){70}{\rule{1.814pt}{0.120pt}}
\multiput(1250.23,721.92)(-57.235,-39.000){2}{\rule{0.907pt}{1.000pt}}
\multiput(1185.12,682.68)(-0.817,-0.498){66}{\rule{1.899pt}{0.120pt}}
\multiput(1189.06,682.92)(-57.059,-37.000){2}{\rule{0.949pt}{1.000pt}}
\multiput(1123.93,645.68)(-0.840,-0.498){64}{\rule{1.944pt}{0.120pt}}
\multiput(1127.96,645.92)(-56.964,-36.000){2}{\rule{0.972pt}{1.000pt}}
\multiput(1062.93,609.68)(-0.840,-0.498){64}{\rule{1.944pt}{0.120pt}}
\multiput(1066.96,609.92)(-56.964,-36.000){2}{\rule{0.972pt}{1.000pt}}
\multiput(1001.64,573.68)(-0.875,-0.497){60}{\rule{2.015pt}{0.120pt}}
\multiput(1005.82,573.92)(-55.818,-34.000){2}{\rule{1.007pt}{1.000pt}}
\multiput(941.05,539.68)(-0.946,-0.497){56}{\rule{2.156pt}{0.120pt}}
\multiput(945.52,539.92)(-56.525,-32.000){2}{\rule{1.078pt}{1.000pt}}
\multiput(879.79,507.68)(-0.977,-0.497){54}{\rule{2.218pt}{0.120pt}}
\multiput(884.40,507.92)(-56.397,-31.000){2}{\rule{1.109pt}{1.000pt}}
\multiput(818.52,476.68)(-1.010,-0.497){52}{\rule{2.283pt}{0.120pt}}
\multiput(823.26,476.92)(-56.261,-30.000){2}{\rule{1.142pt}{1.000pt}}
\multiput(756.92,446.68)(-1.083,-0.497){48}{\rule{2.429pt}{0.120pt}}
\multiput(761.96,446.92)(-55.959,-28.000){2}{\rule{1.214pt}{1.000pt}}
\multiput(695.38,418.68)(-1.148,-0.497){44}{\rule{2.558pt}{0.120pt}}
\multiput(700.69,418.92)(-54.691,-26.000){2}{\rule{1.279pt}{1.000pt}}
\multiput(634.83,392.68)(-1.215,-0.496){42}{\rule{2.690pt}{0.120pt}}
\multiput(640.42,392.92)(-55.417,-25.000){2}{\rule{1.345pt}{1.000pt}}
\multiput(573.72,367.68)(-1.225,-0.493){22}{\rule{2.717pt}{0.119pt}}
\multiput(579.36,367.92)(-31.361,-15.000){2}{\rule{1.358pt}{1.000pt}}
\put(220.0,355.0){\rule[-0.500pt]{79.015pt}{1.000pt}}
\put(220.0,113.0){\rule[-0.500pt]{1.000pt}{58.298pt}}
\put(220.0,113.0){\rule[-0.500pt]{79.015pt}{1.000pt}}
\put(548.0,113.0){\rule[-0.500pt]{1.000pt}{58.298pt}}
\end{picture}
\end{center}
\caption{\label{Fig:bound2} The phase boundaries between the
  $\alpha$-phase, $\beta$-phase and the $0$-phase. The chiral
  symmetry is unbroken in the
  $0$-phase. The $\alpha$-phase and $\beta$-phase break
  the chiral symmetry spontaneously down to a
  flavor symmetry.}
\end{figure}

The effective potential
$V_{eff}$ can be computed. A numerical evaluation of it
in the Euclidean momentum space shows
that the absolute minimum of $V_{eff}(\sigma^2,\phi^2)$ is
located either at
$\sigma^2\ne 0$ and $\phi^2 = 0$, which is the $\alpha$--phase,
or at $\sigma^2 = 0$ and
$\phi^2 \ne 0$, which is called the $\beta$--phase, in
the spontaneous symmetry breaking phases. The phase diagram is
obtained by minimizing $V_{eff}$ with respect to
$\phi^2$ by assuming $\sigma^2=0$ in the $\beta$--phase
and with respect to $\sigma^2$ by assuming $\phi^2=0$ in the
$\alpha$-phase.
The result
is represented in Fig.\ref{Fig:bound2}.
When $\sigma^2 =0$ or $\phi^2=0$
explicit expressions for the effective potential can be derived. They are
\begin{eqnarray}
V_{eff} (\sigma^2,0) &=& {\Lambda^4\over 4\pi}\left \{
{1\over 4}\left ({1\over \alpha_0} - {6\over\pi}\right
){\sigma^2\over
\Lambda^2} + {3\over 2\pi}
ln\left ( 1+{\Lambda^2\over \sigma^2}\right )
{\sigma^4\over\Lambda^4}
- {3\over 2\pi} ln\left (1+{\sigma^2\over
\Lambda^2} \right )\right \},\nonumber \\
&&\label{Vphisig0}
\end{eqnarray}
and
\begin{eqnarray}
V_{eff} (0,\phi^2) &=& {\Lambda^4\over 4\pi}
\left \{
\begin{array}{ll}{\displaystyle
{1\over 4} \left ({1\over\alpha_3} - {4\over\pi}\right )
{\phi^2\over\Lambda^2}+{1\over 6\pi}{\phi^4\over\Lambda^4}}
&
\phi^2\le\Lambda^2\\
{\displaystyle
{1\over 4\alpha_3}{\phi^2\over\Lambda^2} - {1\over 2\pi}
\left ( 1 + 2ln{\phi^2\over\Lambda^2}
 + {2\over 3}{\Lambda^2\over\phi^2}\right )}&
\phi^2>\Lambda^2
\end{array}\right.,\label{EffV4}
\end{eqnarray}
with the reduced dimensionless coupling constants $\alpha_0$ and
$\alpha_3$ defined as
\begin{eqnarray}
\alpha_0 & = & {G_0\Lambda^2\over 4\pi},\label{alf0}\\
\alpha_3 & = & {G_3\Lambda^2\over 8\pi}.\label{alf3}
\end{eqnarray}
The value of $\sigma^2$ that minimize $V_{eff}(\sigma^2,0)$
satisfies
\begin{eqnarray}
{\sigma^2\over\Lambda^2} ln\left ( 1 +
{\Lambda^2\over\sigma^2}
\right ) & = & \left ( 1 - {\pi\over
12\alpha_0}\right ).
\label{alf0min}
\end{eqnarray}
It can not be solved explicitly. The value of $\phi^2$
that minimize
$V_{eff}(0,\phi^2)$ is
\begin{eqnarray}
{\phi^2\over\Lambda^2} &=&\left \{
\begin{array}{ll}\displaystyle
3\left ( 1 - {\pi\over 4\alpha_3}\right
)&{\displaystyle\pi\over 4}\le \alpha_3
\le {\displaystyle 3\pi\over 8}\\
{\displaystyle
2\alpha_3\over \pi}\left ( 1 + \sqrt{1-
{\displaystyle\pi\over 3\alpha_3}}\right
) & \alpha_3>{\displaystyle 3\pi\over 8}
\end{array}\right..\label{alf3min}
\end{eqnarray}
The general form of Eqs.~(\ref{Vphisig0})--(\ref{alf3min}) are
obtained by the
replacement
\begin{eqnarray}
\sigma^2&\to&\sigma^2 + \vec{\pi}^2,\label{sigtosigpi}\\
\phi^2&\to&\bar\phi_{\mu c}\phi^{\mu c} +
\vec{\bar\delta}_{\mu c}\cdot
\vec{\delta}^{\mu c},\label{phitophidel}
\end{eqnarray}
following the symmetry properties of $V_{eff}$. The transition is of
second order across boundaries between
0-phase and the $\alpha$-phase ($\alpha^c_0=
\pi/12$) as well as 0-phase and $\beta$-phase ($\alpha_3=\pi/4$) in
Fig.\ref{Fig:bound2}; it is first order phase transition
across the boundary between $\alpha$- and $\beta$- phase
($\alpha_0\ge \pi/12$ and
$\alpha_3\ge\pi/4$).

The chiral $SU(2)_L\times SU(2)_R$ symmetry is spontaneously broken
down to a $SU(2)_V$ flavor symmetry both in the $\alpha$-phase and in
the $\beta$-phase.

Some of the other properties of the $\beta$-phase are discussed in
more detail in Refs. \cite{YingA}. They are not reproduced here.

\section{Spontaneous separation of baryon number in the $\beta$- and
  $\omega$- phases}
\label{sec:CP-vio}
    The vacuum of the models introduced are studied based on the
assumption that the lowest energy state of the system (vacuum)
contains vanishing baryon number density.
This assumption can be phrased in a different way, namely,
that the baryon number and the antibaryon number in the vacuum of the
system cancel locally leading to net baryon number density zero at
each space-time point. This assumption is not apparent a priori since
it is not an independent one. Whether or not it is true depends, as it
is shown in this study, upon the interaction of the system.
Can the vacuum state of an interacting system contains net baryon
density locally? The answer is yes \cite{YingL,YingB}.

\begin{figure}[h]
\begin{center}
\setlength{\unitlength}{0.240900pt}
\ifx\plotpoint\undefined\newsavebox{\plotpoint}\fi
\sbox{\plotpoint}{\rule[-0.500pt]{1.000pt}{1.000pt}}%
\begin{picture}(1500,900)(0,0)
\font\gnuplot=cmr10 at 10pt
\gnuplot
\sbox{\plotpoint}{\rule[-0.500pt]{1.000pt}{1.000pt}}%
\put(220.0,113.0){\rule[-0.500pt]{4.818pt}{1.000pt}}
\put(198,113){\makebox(0,0)[r]{$-0.80$}}
\put(1416.0,113.0){\rule[-0.500pt]{4.818pt}{1.000pt}}
\put(220.0,266.0){\rule[-0.500pt]{4.818pt}{1.000pt}}
\put(198,266){\makebox(0,0)[r]{$-0.75$}}
\put(1416.0,266.0){\rule[-0.500pt]{4.818pt}{1.000pt}}
\put(220.0,419.0){\rule[-0.500pt]{4.818pt}{1.000pt}}
\put(198,419){\makebox(0,0)[r]{$-0.70$}}
\put(1416.0,419.0){\rule[-0.500pt]{4.818pt}{1.000pt}}
\put(220.0,571.0){\rule[-0.500pt]{4.818pt}{1.000pt}}
\put(198,571){\makebox(0,0)[r]{$-0.65$}}
\put(1416.0,571.0){\rule[-0.500pt]{4.818pt}{1.000pt}}
\put(220.0,724.0){\rule[-0.500pt]{4.818pt}{1.000pt}}
\put(198,724){\makebox(0,0)[r]{$-0.60$}}
\put(1416.0,724.0){\rule[-0.500pt]{4.818pt}{1.000pt}}
\put(220.0,877.0){\rule[-0.500pt]{4.818pt}{1.000pt}}
\put(198,877){\makebox(0,0)[r]{$-0.55$}}
\put(1416.0,877.0){\rule[-0.500pt]{4.818pt}{1.000pt}}
\put(220.0,113.0){\rule[-0.500pt]{1.000pt}{4.818pt}}
\put(220,68){\makebox(0,0){$-0.5$}}
\put(220.0,857.0){\rule[-0.500pt]{1.000pt}{4.818pt}}
\put(342.0,113.0){\rule[-0.500pt]{1.000pt}{4.818pt}}
\put(342,68){\makebox(0,0){$-0.4$}}
\put(342.0,857.0){\rule[-0.500pt]{1.000pt}{4.818pt}}
\put(463.0,113.0){\rule[-0.500pt]{1.000pt}{4.818pt}}
\put(463,68){\makebox(0,0){$-0.3$}}
\put(463.0,857.0){\rule[-0.500pt]{1.000pt}{4.818pt}}
\put(585.0,113.0){\rule[-0.500pt]{1.000pt}{4.818pt}}
\put(585,68){\makebox(0,0){$-0.2$}}
\put(585.0,857.0){\rule[-0.500pt]{1.000pt}{4.818pt}}
\put(706.0,113.0){\rule[-0.500pt]{1.000pt}{4.818pt}}
\put(706,68){\makebox(0,0){$-0.1$}}
\put(706.0,857.0){\rule[-0.500pt]{1.000pt}{4.818pt}}
\put(828.0,113.0){\rule[-0.500pt]{1.000pt}{4.818pt}}
\put(828,68){\makebox(0,0){$0$}}
\put(828.0,857.0){\rule[-0.500pt]{1.000pt}{4.818pt}}
\put(950.0,113.0){\rule[-0.500pt]{1.000pt}{4.818pt}}
\put(950,68){\makebox(0,0){$0.1$}}
\put(950.0,857.0){\rule[-0.500pt]{1.000pt}{4.818pt}}
\put(1071.0,113.0){\rule[-0.500pt]{1.000pt}{4.818pt}}
\put(1071,68){\makebox(0,0){$0.2$}}
\put(1071.0,857.0){\rule[-0.500pt]{1.000pt}{4.818pt}}
\put(1193.0,113.0){\rule[-0.500pt]{1.000pt}{4.818pt}}
\put(1193,68){\makebox(0,0){$0.3$}}
\put(1193.0,857.0){\rule[-0.500pt]{1.000pt}{4.818pt}}
\put(1314.0,113.0){\rule[-0.500pt]{1.000pt}{4.818pt}}
\put(1314,68){\makebox(0,0){$0.4$}}
\put(1314.0,857.0){\rule[-0.500pt]{1.000pt}{4.818pt}}
\put(1436.0,113.0){\rule[-0.500pt]{1.000pt}{4.818pt}}
\put(1436,68){\makebox(0,0){$0.5$}}
\put(1436.0,857.0){\rule[-0.500pt]{1.000pt}{4.818pt}}
\put(220.0,113.0){\rule[-0.500pt]{292.934pt}{1.000pt}}
\put(1436.0,113.0){\rule[-0.500pt]{1.000pt}{184.048pt}}
\put(220.0,877.0){\rule[-0.500pt]{292.934pt}{1.000pt}}
\put(45,495){\makebox(0,0){$v_{eff}$}}
\put(828,23){\makebox(0,0){$x$}}
\put(828,266){\makebox(0,0){$\alpha_{3'}=1.0$}}
\put(269,923){\makebox(0,0){$\times 10^{-2}$}}
\put(220.0,113.0){\rule[-0.500pt]{1.000pt}{184.048pt}}
\sbox{\plotpoint}{\rule[-0.175pt]{0.350pt}{0.350pt}}%
\put(317,638){\usebox{\plotpoint}}
\multiput(317.48,628.92)(0.502,-3.133){23}{\rule{0.121pt}{2.188pt}}
\multiput(316.27,633.46)(13.000,-73.460){2}{\rule{0.350pt}{1.094pt}}
\multiput(330.48,551.93)(0.502,-2.769){23}{\rule{0.121pt}{1.945pt}}
\multiput(329.27,555.96)(13.000,-64.963){2}{\rule{0.350pt}{0.973pt}}
\multiput(343.48,483.93)(0.502,-2.406){23}{\rule{0.121pt}{1.703pt}}
\multiput(342.27,487.47)(13.000,-56.466){2}{\rule{0.350pt}{0.851pt}}
\multiput(356.48,424.58)(0.502,-2.179){21}{\rule{0.121pt}{1.546pt}}
\multiput(355.27,427.79)(12.000,-46.792){2}{\rule{0.350pt}{0.773pt}}
\multiput(368.48,375.83)(0.502,-1.720){23}{\rule{0.121pt}{1.245pt}}
\multiput(367.27,378.42)(13.000,-40.416){2}{\rule{0.350pt}{0.623pt}}
\multiput(381.48,333.73)(0.502,-1.397){23}{\rule{0.121pt}{1.030pt}}
\multiput(380.27,335.86)(13.000,-32.863){2}{\rule{0.350pt}{0.515pt}}
\multiput(394.48,299.51)(0.502,-1.114){23}{\rule{0.121pt}{0.841pt}}
\multiput(393.27,301.25)(13.000,-26.254){2}{\rule{0.350pt}{0.421pt}}
\multiput(407.48,271.97)(0.502,-0.949){21}{\rule{0.121pt}{0.729pt}}
\multiput(406.27,273.49)(12.000,-20.487){2}{\rule{0.350pt}{0.365pt}}
\multiput(419.48,250.96)(0.502,-0.589){23}{\rule{0.121pt}{0.491pt}}
\multiput(418.27,251.98)(13.000,-13.980){2}{\rule{0.350pt}{0.246pt}}
\multiput(432.00,237.02)(0.672,-0.503){17}{\rule{0.542pt}{0.121pt}}
\multiput(432.00,237.27)(11.874,-10.000){2}{\rule{0.271pt}{0.350pt}}
\multiput(445.00,227.02)(1.186,-0.505){9}{\rule{0.846pt}{0.122pt}}
\multiput(445.00,227.27)(11.244,-6.000){2}{\rule{0.423pt}{0.350pt}}
\multiput(470.00,222.47)(3.170,0.516){3}{\rule{1.604pt}{0.124pt}}
\multiput(470.00,221.27)(9.670,3.000){2}{\rule{0.802pt}{0.350pt}}
\multiput(483.00,225.47)(0.856,0.504){13}{\rule{0.656pt}{0.121pt}}
\multiput(483.00,224.27)(11.638,8.000){2}{\rule{0.328pt}{0.350pt}}
\multiput(496.00,233.48)(0.672,0.503){17}{\rule{0.542pt}{0.121pt}}
\multiput(496.00,232.27)(11.874,10.000){2}{\rule{0.271pt}{0.350pt}}
\multiput(509.00,243.48)(0.508,0.502){23}{\rule{0.438pt}{0.121pt}}
\multiput(509.00,242.27)(12.092,13.000){2}{\rule{0.219pt}{0.350pt}}
\multiput(522.48,256.00)(0.502,0.685){21}{\rule{0.121pt}{0.554pt}}
\multiput(521.27,256.00)(12.000,14.850){2}{\rule{0.350pt}{0.277pt}}
\multiput(534.48,272.00)(0.502,0.670){23}{\rule{0.121pt}{0.545pt}}
\multiput(533.27,272.00)(13.000,15.868){2}{\rule{0.350pt}{0.273pt}}
\multiput(547.48,289.00)(0.502,0.791){23}{\rule{0.121pt}{0.626pt}}
\multiput(546.27,289.00)(13.000,18.701){2}{\rule{0.350pt}{0.313pt}}
\multiput(560.48,309.00)(0.502,0.791){23}{\rule{0.121pt}{0.626pt}}
\multiput(559.27,309.00)(13.000,18.701){2}{\rule{0.350pt}{0.313pt}}
\multiput(573.48,329.00)(0.502,0.949){21}{\rule{0.121pt}{0.729pt}}
\multiput(572.27,329.00)(12.000,20.487){2}{\rule{0.350pt}{0.365pt}}
\multiput(585.48,351.00)(0.502,0.912){23}{\rule{0.121pt}{0.707pt}}
\multiput(584.27,351.00)(13.000,21.533){2}{\rule{0.350pt}{0.353pt}}
\multiput(598.48,374.00)(0.502,0.912){23}{\rule{0.121pt}{0.707pt}}
\multiput(597.27,374.00)(13.000,21.533){2}{\rule{0.350pt}{0.353pt}}
\multiput(611.48,397.00)(0.502,0.912){23}{\rule{0.121pt}{0.707pt}}
\multiput(610.27,397.00)(13.000,21.533){2}{\rule{0.350pt}{0.353pt}}
\multiput(624.48,420.00)(0.502,0.993){21}{\rule{0.121pt}{0.758pt}}
\multiput(623.27,420.00)(12.000,21.426){2}{\rule{0.350pt}{0.379pt}}
\multiput(636.48,443.00)(0.502,0.912){23}{\rule{0.121pt}{0.707pt}}
\multiput(635.27,443.00)(13.000,21.533){2}{\rule{0.350pt}{0.353pt}}
\multiput(649.48,466.00)(0.502,0.872){23}{\rule{0.121pt}{0.680pt}}
\multiput(648.27,466.00)(13.000,20.589){2}{\rule{0.350pt}{0.340pt}}
\multiput(662.48,488.00)(0.502,0.831){23}{\rule{0.121pt}{0.653pt}}
\multiput(661.27,488.00)(13.000,19.645){2}{\rule{0.350pt}{0.326pt}}
\multiput(675.48,509.00)(0.502,0.831){23}{\rule{0.121pt}{0.653pt}}
\multiput(674.27,509.00)(13.000,19.645){2}{\rule{0.350pt}{0.326pt}}
\multiput(688.48,530.00)(0.502,0.817){21}{\rule{0.121pt}{0.642pt}}
\multiput(687.27,530.00)(12.000,17.668){2}{\rule{0.350pt}{0.321pt}}
\multiput(700.48,549.00)(0.502,0.710){23}{\rule{0.121pt}{0.572pt}}
\multiput(699.27,549.00)(13.000,16.813){2}{\rule{0.350pt}{0.286pt}}
\multiput(713.48,567.00)(0.502,0.629){23}{\rule{0.121pt}{0.518pt}}
\multiput(712.27,567.00)(13.000,14.924){2}{\rule{0.350pt}{0.259pt}}
\multiput(726.48,583.00)(0.502,0.589){23}{\rule{0.121pt}{0.491pt}}
\multiput(725.27,583.00)(13.000,13.980){2}{\rule{0.350pt}{0.246pt}}
\multiput(739.48,598.00)(0.502,0.553){21}{\rule{0.121pt}{0.467pt}}
\multiput(738.27,598.00)(12.000,12.031){2}{\rule{0.350pt}{0.233pt}}
\multiput(751.00,611.48)(0.607,0.502){19}{\rule{0.501pt}{0.121pt}}
\multiput(751.00,610.27)(11.960,11.000){2}{\rule{0.251pt}{0.350pt}}
\multiput(764.00,622.47)(0.753,0.503){15}{\rule{0.593pt}{0.121pt}}
\multiput(764.00,621.27)(11.769,9.000){2}{\rule{0.297pt}{0.350pt}}
\multiput(777.00,631.47)(0.856,0.504){13}{\rule{0.656pt}{0.121pt}}
\multiput(777.00,630.27)(11.638,8.000){2}{\rule{0.328pt}{0.350pt}}
\multiput(790.00,639.47)(1.358,0.507){7}{\rule{0.928pt}{0.122pt}}
\multiput(790.00,638.27)(10.075,5.000){2}{\rule{0.464pt}{0.350pt}}
\multiput(802.00,644.47)(3.170,0.516){3}{\rule{1.604pt}{0.124pt}}
\multiput(802.00,643.27)(9.670,3.000){2}{\rule{0.802pt}{0.350pt}}
\put(815,646.77){\rule{3.132pt}{0.350pt}}
\multiput(815.00,646.27)(6.500,1.000){2}{\rule{1.566pt}{0.350pt}}
\put(828,646.77){\rule{3.132pt}{0.350pt}}
\multiput(828.00,647.27)(6.500,-1.000){2}{\rule{1.566pt}{0.350pt}}
\multiput(841.00,646.02)(3.170,-0.516){3}{\rule{1.604pt}{0.124pt}}
\multiput(841.00,646.27)(9.670,-3.000){2}{\rule{0.802pt}{0.350pt}}
\multiput(854.00,643.02)(1.358,-0.507){7}{\rule{0.928pt}{0.122pt}}
\multiput(854.00,643.27)(10.075,-5.000){2}{\rule{0.464pt}{0.350pt}}
\multiput(866.00,638.02)(0.856,-0.504){13}{\rule{0.656pt}{0.121pt}}
\multiput(866.00,638.27)(11.638,-8.000){2}{\rule{0.328pt}{0.350pt}}
\multiput(879.00,630.02)(0.753,-0.503){15}{\rule{0.593pt}{0.121pt}}
\multiput(879.00,630.27)(11.769,-9.000){2}{\rule{0.297pt}{0.350pt}}
\multiput(892.00,621.02)(0.607,-0.502){19}{\rule{0.501pt}{0.121pt}}
\multiput(892.00,621.27)(11.960,-11.000){2}{\rule{0.251pt}{0.350pt}}
\multiput(905.48,609.06)(0.502,-0.553){21}{\rule{0.121pt}{0.467pt}}
\multiput(904.27,610.03)(12.000,-12.031){2}{\rule{0.350pt}{0.233pt}}
\multiput(917.48,595.96)(0.502,-0.589){23}{\rule{0.121pt}{0.491pt}}
\multiput(916.27,596.98)(13.000,-13.980){2}{\rule{0.350pt}{0.246pt}}
\multiput(930.48,580.85)(0.502,-0.629){23}{\rule{0.121pt}{0.518pt}}
\multiput(929.27,581.92)(13.000,-14.924){2}{\rule{0.350pt}{0.259pt}}
\multiput(943.48,564.63)(0.502,-0.710){23}{\rule{0.121pt}{0.572pt}}
\multiput(942.27,565.81)(13.000,-16.813){2}{\rule{0.350pt}{0.286pt}}
\multiput(956.48,546.34)(0.502,-0.817){21}{\rule{0.121pt}{0.642pt}}
\multiput(955.27,547.67)(12.000,-17.668){2}{\rule{0.350pt}{0.321pt}}
\multiput(968.48,527.29)(0.502,-0.831){23}{\rule{0.121pt}{0.653pt}}
\multiput(967.27,528.64)(13.000,-19.645){2}{\rule{0.350pt}{0.326pt}}
\multiput(981.48,506.29)(0.502,-0.831){23}{\rule{0.121pt}{0.653pt}}
\multiput(980.27,507.64)(13.000,-19.645){2}{\rule{0.350pt}{0.326pt}}
\multiput(994.48,485.18)(0.502,-0.872){23}{\rule{0.121pt}{0.680pt}}
\multiput(993.27,486.59)(13.000,-20.589){2}{\rule{0.350pt}{0.340pt}}
\multiput(1007.48,463.07)(0.502,-0.912){23}{\rule{0.121pt}{0.707pt}}
\multiput(1006.27,464.53)(13.000,-21.533){2}{\rule{0.350pt}{0.353pt}}
\multiput(1020.48,439.85)(0.502,-0.993){21}{\rule{0.121pt}{0.758pt}}
\multiput(1019.27,441.43)(12.000,-21.426){2}{\rule{0.350pt}{0.379pt}}
\multiput(1032.48,417.07)(0.502,-0.912){23}{\rule{0.121pt}{0.707pt}}
\multiput(1031.27,418.53)(13.000,-21.533){2}{\rule{0.350pt}{0.353pt}}
\multiput(1045.48,394.07)(0.502,-0.912){23}{\rule{0.121pt}{0.707pt}}
\multiput(1044.27,395.53)(13.000,-21.533){2}{\rule{0.350pt}{0.353pt}}
\multiput(1058.48,371.07)(0.502,-0.912){23}{\rule{0.121pt}{0.707pt}}
\multiput(1057.27,372.53)(13.000,-21.533){2}{\rule{0.350pt}{0.353pt}}
\multiput(1071.48,347.97)(0.502,-0.949){21}{\rule{0.121pt}{0.729pt}}
\multiput(1070.27,349.49)(12.000,-20.487){2}{\rule{0.350pt}{0.365pt}}
\multiput(1083.48,326.40)(0.502,-0.791){23}{\rule{0.121pt}{0.626pt}}
\multiput(1082.27,327.70)(13.000,-18.701){2}{\rule{0.350pt}{0.313pt}}
\multiput(1096.48,306.40)(0.502,-0.791){23}{\rule{0.121pt}{0.626pt}}
\multiput(1095.27,307.70)(13.000,-18.701){2}{\rule{0.350pt}{0.313pt}}
\multiput(1109.48,286.74)(0.502,-0.670){23}{\rule{0.121pt}{0.545pt}}
\multiput(1108.27,287.87)(13.000,-15.868){2}{\rule{0.350pt}{0.273pt}}
\multiput(1122.48,269.70)(0.502,-0.685){21}{\rule{0.121pt}{0.554pt}}
\multiput(1121.27,270.85)(12.000,-14.850){2}{\rule{0.350pt}{0.277pt}}
\multiput(1134.00,255.02)(0.508,-0.502){23}{\rule{0.438pt}{0.121pt}}
\multiput(1134.00,255.27)(12.092,-13.000){2}{\rule{0.219pt}{0.350pt}}
\multiput(1147.00,242.02)(0.672,-0.503){17}{\rule{0.542pt}{0.121pt}}
\multiput(1147.00,242.27)(11.874,-10.000){2}{\rule{0.271pt}{0.350pt}}
\multiput(1160.00,232.02)(0.856,-0.504){13}{\rule{0.656pt}{0.121pt}}
\multiput(1160.00,232.27)(11.638,-8.000){2}{\rule{0.328pt}{0.350pt}}
\multiput(1173.00,224.02)(3.170,-0.516){3}{\rule{1.604pt}{0.124pt}}
\multiput(1173.00,224.27)(9.670,-3.000){2}{\rule{0.802pt}{0.350pt}}
\put(458.0,222.0){\rule[-0.175pt]{2.891pt}{0.350pt}}
\multiput(1198.00,222.47)(1.186,0.505){9}{\rule{0.846pt}{0.122pt}}
\multiput(1198.00,221.27)(11.244,6.000){2}{\rule{0.423pt}{0.350pt}}
\multiput(1211.00,228.48)(0.672,0.503){17}{\rule{0.542pt}{0.121pt}}
\multiput(1211.00,227.27)(11.874,10.000){2}{\rule{0.271pt}{0.350pt}}
\multiput(1224.48,238.00)(0.502,0.589){23}{\rule{0.121pt}{0.491pt}}
\multiput(1223.27,238.00)(13.000,13.980){2}{\rule{0.350pt}{0.246pt}}
\multiput(1237.48,253.00)(0.502,0.949){21}{\rule{0.121pt}{0.729pt}}
\multiput(1236.27,253.00)(12.000,20.487){2}{\rule{0.350pt}{0.365pt}}
\multiput(1249.48,275.00)(0.502,1.114){23}{\rule{0.121pt}{0.841pt}}
\multiput(1248.27,275.00)(13.000,26.254){2}{\rule{0.350pt}{0.421pt}}
\multiput(1262.48,303.00)(0.502,1.397){23}{\rule{0.121pt}{1.030pt}}
\multiput(1261.27,303.00)(13.000,32.863){2}{\rule{0.350pt}{0.515pt}}
\multiput(1275.48,338.00)(0.502,1.720){23}{\rule{0.121pt}{1.245pt}}
\multiput(1274.27,338.00)(13.000,40.416){2}{\rule{0.350pt}{0.623pt}}
\multiput(1288.48,381.00)(0.502,2.179){21}{\rule{0.121pt}{1.546pt}}
\multiput(1287.27,381.00)(12.000,46.792){2}{\rule{0.350pt}{0.773pt}}
\multiput(1300.48,431.00)(0.502,2.406){23}{\rule{0.121pt}{1.703pt}}
\multiput(1299.27,431.00)(13.000,56.466){2}{\rule{0.350pt}{0.851pt}}
\multiput(1313.48,491.00)(0.502,2.769){23}{\rule{0.121pt}{1.945pt}}
\multiput(1312.27,491.00)(13.000,64.963){2}{\rule{0.350pt}{0.973pt}}
\put(1186.0,222.0){\rule[-0.175pt]{2.891pt}{0.350pt}}
\end{picture}
\end{center}
\caption{\label{Fig:Baryon}
         The effective potential $v_{eff}=V_{eff}/\Lambda^4$
         as a function of $x=\mu/\Lambda$.}
\end{figure}
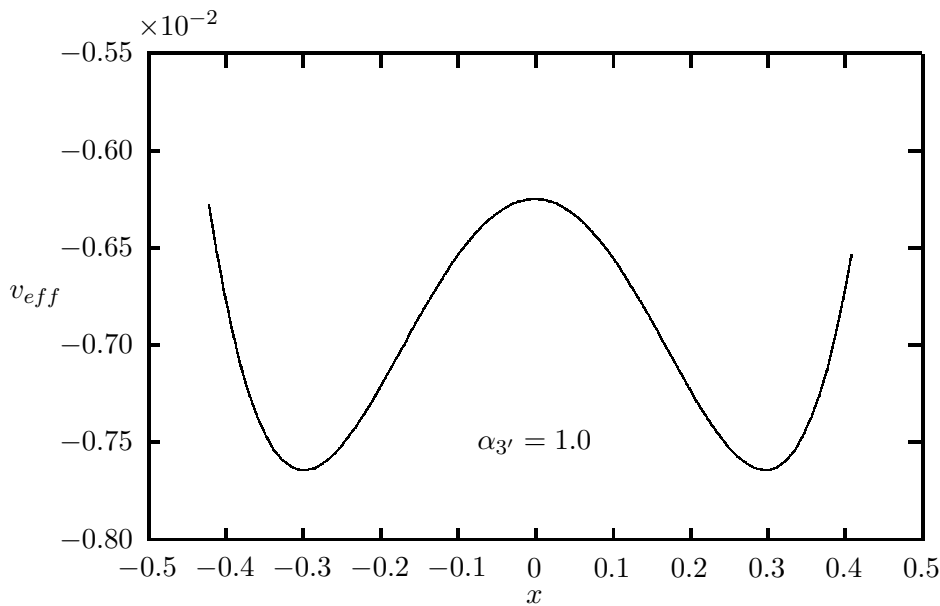

  To investigate this question, a Lagrangian density $\tilde {\cal L}$
differ from the
original one Eq. \ref{Lagrangian} by an additional $\mu_\alpha
j^\alpha_B$ term, with $\mu^\alpha$ a statistical gauge field
\cite{YingB} and $j^\alpha_B$ the baryon number current density, can be
used \cite{YingL}. In the 8-component
representation for the Dirac spinor, it takes the
following form
\begin{eqnarray}
      \tilde {\cal L}&=&
       {1\over 2} \bar\Psi\left ( i\rlap\slash\partial
              + O_3 \rlap\slash\mu \right ) \Psi + {\cal L}_{int}.
\label{Lagrangian2}
\end{eqnarray}
In the equilibrium statistical mechanics formulated in a path
integration Language,
such an additional term is the one needed in a grand-canonical assemble.
For the $\beta$- and $\omega$- phases,
the question of what's the configuration for
$\mu^\alpha$ in the  lowest energy state of the system,
which is the one corresponding to the
vacuum by definition, can be studied using this Lagrangian
density in the conventional formalism treated in the Euclidean space
\footnote{For a consistent discussion of the $\alpha$-, $\beta$- and
  $\omega$- phases, certain new computation method has to be
  introduced. They are given in Ref. \cite{YingB}.}. The
resulting effective potential for the $\omega$-phase as a function of
$\mu\equiv \sqrt{\mu^2}/\Lambda$ with a definite $\alpha_{3'}$ (and
therefore $\chi^2$) is shown in Fig. \ref{Fig:Baryon}. It can be
seen that the values for $\mu$ that minimize the effective
potential are nonzero; this is somewhat counterintuitive. For the
$\beta$-phase, the result is similar \cite{YingL}.
It is shown in Ref. \cite{YingB} that in the $\alpha$-phase, the lowest energy
configuration for $\mu$ is located at $\mu=0$.

A nonvanishing $\mu$ implies non-vanishing local baryon number density
in the vacuum of the system. It means that {\em baryon rich and antibaryon
rich regions can be generated spontaneously in the $\beta$- and
$\omega$- phases of the systems considered}. A nonvanishing $\mu$ also
spontaneously breaks the CP invariance of the original system.
These properties of the $\omega$- and $\beta$- phases, in which the
$U(1)$ phase symmetry corresponding to the baryon number conservation
is spontaneously broken down,
have the potential of solving \cite{YingL} the old
problem \cite{Steigman} in the
idea of the matter--antimatter symmetric universe
\cite{Omnes} based on standard big--bang cosmology.

The above mentioned results are actually not difficulty to perceive. In the
$\alpha$-phase, correlated
quark--antiquark pairs condense in the vacuum, they can not be
spontaneously separated. In the $\beta$- and $\omega$-phases,
correlated quark--quark pairs and antiquark--antiquark pairs condense;
it is conceivable that they prefer to separate locally while keeping
the global net baryon number unchanged. This qualitative picture can actually
be
more rigorously substantiated by using field theoretical method.
The details are given in Ref. \cite{YingB}.


\part{ Traces of $\beta$- and $\omega$- Phases in Nuclear Systems}

     Four possible phases of the light quark system are discovered in
Part I of this report. Since the models introduced are not derived
from the QCD Lagrangian density, their relevancy can be established
either by deducing them from the fundamental theory (QCD) or by
checking whether or not the consequences of these models are
observable in nuclear systems, like, the static case of inside a
nucleon, the dynamical processes involving hadronic reactions, etc.

   Deducing the quark--quark interaction terms are not the goal of
this research; it is certainly an nontrivial and interesting topic
to be studied in the future. The relevancy of the results of the
first part of the report are assessed by looking for the possible
existence of the phases found there in realistic nuclear systems.
Two kinds of systems are considered: 1) inside a nucleon and 2) in the
high energy $e^+e^-$ annihilation processes.

    The large scale vacuum phase for the strong interaction vacuum is
in the $\alpha$-phase. There are large body of empirical evidences
supporting such a notion. The reason behind the search for the $\beta$- or
$\omega$- phases inside a nucleon and in the hadronization processes
of the $e^+e^-$ reaction is inspired by the facts that sufficient
high baryon density can cause the strong interaction vacuum to flip
into $\beta$- or $\omega$- phase \cite{YingL,YingB}. There might be
sufficiently high baryon density inside a nucleon and in the
hadronization process of the $e^+e^-$ reaction.

    The results derived in part I are for a uniform system with
its spatial and temporal extension going to infinity. When
 these results are used in case of a nucleon which is of
finite extension in space or in case of an $e^+e^-$ annihilation
reaction which is finite both in time duraction and spatial extension,
fluctuations effects has to be taken into account. Since what is
seeking in this work is the null effects, which means that the effects
are absent without the existence of the $\beta$- or $\omega$- phase, we
belief that the question of existence of these phases are not affected
by the effects of the fluctuations and a mean field picture is
still a good starting point for systematical improvement of the
result for technical reasons. This will be elaborated in more
detail in the following discussions, which do not depend on mean field
pictures.

\chapter{Extended PCAC Relation for a Nucleon}

\section{A QCD chiral Ward identity and PCAC relation}

  The QCD Lagrangian density
\begin{eqnarray}
      {\cal L} &=& -{1\over 8} Tr G^{\mu\nu} G_{\mu\nu} +
                     \bar\psi (i\rlap\slash\hskip -2.5pt D - m_0)\psi,
\label{QCD-Lag}
\end{eqnarray}
with $G^{\mu\nu}$ the gluon field strength tensor, $\psi$ the quark field,
has a global chiral $SU(2)_L\times SU(2)_R$ symmetry if $m_0$ is set to zero.
There is a QCD Ward identity for the divergence of $A^a_\mu$
\begin{eqnarray}
\partial^\mu A^a_\mu &=& 2 m_0 \bar\psi i\gamma^5 \tau^a \psi
\label{QCD-WI}
\end{eqnarray}
in the $m_0\ne 0$ case at the ``classical'' level.
Here $\tau^a$ (a=1,2,3) is one of the Pauli matrices in the isospin space.
Unlike its $U(1)_L\times U(1)_R$ chiral symmetry, which is anomalous,
the $SU(2)_L\times SU(2)_R$ chiral symmetry of QCD is anomalies free.
Therefore Eq. \ref{QCD-WI} continues to be valid when the system gets
quantized, which turns it into an operator equation. It has, however,
no predictive power in this form since the matrix elements of the
right hand side (r.h.s.) of it between physical hadronic states
is not immediately known.

The PCAC relation states that
\begin{eqnarray}
\partial^\mu A^a_\mu &=& m_\pi^2 f_\pi \phi_\pi^a,
\label{PCAC-rel1}
\end{eqnarray}
While QCD chiral Ward identity follows directly from the QCD
Lagrangian, the PCAC relationship is an empirically law, which agrees
with data well when $q^2\sim m^2_\pi \sim 0$. It implies the following
weak equality
\begin{eqnarray}
    \phi_\pi^a &\sim& \bar\psi i\gamma^5 \tau^a \psi
\label{PCAC-QCD}
\end{eqnarray}
for their matrix elements in the momentum transfer regions considered.

{\em
How does one go from the QCD chiral Ward idnetity Eq. \ref{QCD-WI} to
the PCAC relation Eq. \ref{PCAC-rel1}?
How good is it when $q^2$ is allowed to go
away from the pion mass shell?
Or, how much difference does one expect between
$ \partial^\mu A^a_\mu$ and $
f_\pi m^2_\pi \phi_\pi$ when their
matrix elements between single nucleon states are taken?
} These questions are studied in this work.

An necessary condition for it to be satified is to has its matrix
elements between physical hadronic states satisfy the same equation.
The PCAC relation in the mesonic sector, namely its matrix elements
between meson states, is known to be satisfied. Is it satisfied
in the baryonic sector? A closer look at it is necessary in order to
have an answer to it.

\section{PCAC relation for nucleon states}

The matrix elements of $\partial^\mu A_\mu^a
-f_\pi m_\pi^2 \phi_\pi$ between single nucleon states are
parameterized as
\begin{eqnarray}
\mathopen{\langle p'\,|} \partial^\mu A^a_\mu(0) - f_\pi m^2_\pi
\phi_\pi(0)
\mathclose{|\, p\rangle} = m^2_\pi C(q^2)
 \bar U(p')i\gamma^5\tau^a U(p)
\end{eqnarray}
with $C(q^2)$ a measure of the error of the PCAC relationship.
In terms of various nucleon invariant form factors, it can be written as
\begin{eqnarray}
q^2 \left [ m g_A + (q^2-m^2_\pi) g_P/2 \right ]
+  m_\pi^2 \left [g_{\pi NN} f_\pi - m g_A \right ]
& = & m^2_\pi (q^2-m^2_\pi) C(q^2).
\label{PCAC-rel2}
\end{eqnarray}
Here $m$ is the mass of a nucleon. $g_A$, $g_{\pi NN}$, $f_\pi$ and $g_P$ are
nucleon axial vector current form factor, the pion-nucleon coupling
constant, pion decay constant and the nucleon pseudoscalar form factor
respectively. They are functions of both $q^2$ and $m^2_\pi$.

Let's define two functions
\begin{eqnarray}
 A(q^2,m^2_\pi)& =& 2 m g_A + (q^2-m^2_\pi) g_P\\
\label{A-func}
 B(q^2,m^2_\pi) &=& g_{\pi NN} f_\pi - m g_A,
\label{B-func}
\end{eqnarray}
where the $m^2_\pi$ dependence is written explicitly.
It follows from Eq. \ref{PCAC-rel2} that
\begin{eqnarray}
A(q^2,m^2_\pi) = 2 m^2_\pi C(q^2,m^2_\pi),\\
\label{A-func2}
B(q^2,m^2_\pi) = -m^2_\pi C(q^2,m^2_\pi),
\label{B-func2}
\end{eqnarray}
by noting that $A(q^2,m^2_\pi)$, $B(q^2,m^2_\pi)$ and
$C(q^2,m^2_\pi)$ are slow varying functions of $q^2$ so that the
coefficient of the explicit $q^2$ dependent term
in Eq. \ref{PCAC-rel2} should vanish (for a more detailed analysis
of this assumption, see Ref. \cite{Epcac}).

\section{Chiral symmetry and two basics relations between nucleon form
        factors}

It can be noted that in the chiral symmetry limit $m^2_\pi\to 0$
\begin{eqnarray}
A(q^2,0) &=& \lim_{m^2_\pi\to 0} [2 m g_A + (q^2-m^2_\pi) g_P] = 0,
\label{Aq20}
\end{eqnarray}
which represents the conservation of the axial vector current in that
limit. It is also expected that
\begin{eqnarray}
\lim_{q^2\to\infty} A(q^2,m^2_\pi)&=&0,
\label{Aq2infty}
\end{eqnarray}
which means that the effects of $m_0$ in the QCD
Lagrangian can be neglected when the momentum transfer $q^2>>m^2_\pi$.
In the region $q^2\sim m^2_\pi$, $A(q^2,m^2_\pi) = m^2_\pi C(q^2
,m^2_\pi)$, which is small as it will be shown in the following. So,
our first basic assumption is
\begin{eqnarray}
A(q^2,m^2_\pi) &\approx& A(q^2,0)=0,
\end{eqnarray}
which leads, in the realistic case of $m_0\ne 0$, to the the following
equation
\begin{eqnarray}
 m g_A(q^2)+ {\displaystyle 1\over\displaystyle 2}(q^2-m^2_\pi)g_P(q^2)
  & = & m^2_\pi C(q^2,m^2_\pi)\approx 0.
\label{Relation1}
\end{eqnarray}
The correction to it is of order $O(m^2_\pi/M^2_A)\sim 1\%$ ($M_A$ is
the lightest meson next to pion in the pionic channel).
This equation is supported by recent experimental measurements
\cite{GaGp1,GaGp2} within $0<-q^2<0.2$ GeV$^2$. The agreement of the above
relationship with experiments is quite good. It is important further
investigation of the relationship between $g_A(q^2)$ and $g_P(q^2)$
can be carried out.

Here, we shall assume Eq. \ref{Relation1} to be true (within an error of order
$m^2_\pi/M^2\sim m_0/M \sim 1-2 \%$). Then
\begin{eqnarray}
    g_{\pi NN}(q^2) f_\pi(q^2) - m g_A(q^2) &=& -m^2_\pi
      C(q^2,m^2_\pi)\approx 0.
\label{Relation2}
\end{eqnarray}
The correction to it is also of order $1\%$. It is indeed the case
on the pion mass shell $q^2=m^2_\pi$ where
the value of $g_A(m^2_\pi)$, $g_{\pi NN}(m^2_\pi)$ and
$f_\pi(m^2_\pi)$ can be deduced from experimental data. Data
suggests that $m^2_\pi C(m^2_\pi,m^2_\pi)\sim 1-2\%$.

\section{The determination of valid $q^2$ region}

Let's evaluate the $q^2$ dependence of $C(q^2,m^2_\pi)$ so that the range of
validity of Eqs. \ref{Relation1} and \ref{Relation2}
can be assessed by using
an once subtracted sum rule, namely
\begin{eqnarray}
 C(q^2) &=& C(m^2_\pi) + {q^2-m^2_\pi\over\pi}\int^\infty_{s_{th}}
             {Im C(s)\over (s-q^2)(s-m^2_\pi)},
\label{C-eval}
\end{eqnarray}
where the $m^2_\pi$ dependence of $C(q^2,m^2_\pi)$ is suppressed for
simpicity. Since at small $|q^2|$, the details of $ImC(s)$ is not important,
we can use a step function to make an estimate, namely,
$ImC(s)\approx \alpha \theta(s-s_{th})$, with $m^2_\pi\alpha \sim 1-2\%$.
In this simplified case
\begin{eqnarray}
  C(q^2) & = & C(m^2_\pi) + {\alpha\over\pi} ln \left (
          {s_{th}-q^2\over s_{th} - m_\pi^2}\right),
\label{Cq2-1}
\end{eqnarray}
where the $m^2_\pi$ dependence of $C(q^2,m^2_\pi)$ is suppressed for
simpicity. Since at small $|q^2|$, the details of $ImC(s)$ is not important,
we can use a step function to make an estimate, namely,
$ImC(s)\approx \alpha \theta(s-s_{th})$, with $m^2_\pi\alpha \sim 1-2\%$.
In this simplified case
\begin{eqnarray}
  C(q^2) &=& C(m^2_\pi) + {\alpha\over\pi} ln \left (
          {s_{th}-q^2\over s_{th} - m_\pi^2}\right).
\label{Cq2-2}
\end{eqnarray}
In order for the corrections to our two basic equations Eqs. \ref{Relation1}
and
\ref{Relation2} to increase by another $1\%$,
$|C(q^2)-C(m^2_\pi)|$ has to increase by $100\%$,
which means that the region of validity of Eqs. \ref{Relation1} and
\ref{Relation2} is
$q^2< 0.7s_{th}$. The range of validity of these two equations in the
negative $q^2$ region is much smaller than --0.7$s_{th}$.

The next step is to estimate the value of $s_{th}$. The lowest
value of $s_{th}$ in the pionic channel is below $9 m_\pi^2$, which
corresponds to an anomalous threshold for the three pion state. However, the
effective $s_{th}$ correspond to that of the $\rho\pi$ threshold,
which is considerablly larger than $9 m_\pi^2$. This is a consequence of
the fact that the underlying dynamics of QCD is chiral invariant
except for a small mass term. From this fact the dynamical (operator)
equation or QCD chiral Ward identity
follows. This dynamical equation ensures that
the operator $\bar\Psi i\gamma^5 \tau^a O_3\Psi$ can only excite a
longitudinal vector excitation since it is proportional to the
divergence of an axial vector operator field. Therefore the state in which
the three pions are all in a s-state is dynamically forbiden. The allowed
state which dominates the dispersion relation is the one where
two of the three pions have a relative angular momentum  of 1, which
is itself dominated by the $\rho$ excitation strength. Therefore $s_{th}\sim
(m_\rho+m_\pi)^2$ and the range of $q^2$ in which Eqs. \ref{Relation1}
and \ref{Relation2}  are valid is 30 to 35 times larger than $m_\pi^2 \approx
0.02 GeV^2$.

\section{Test of PCAC relation for a nucleon}

Eq. \ref{Relation2} gives a specific relation between $g_A(q^2)$ and
$g_{\pi NN}(q^2)f_\pi(q^2)$.
We study whether or not it is consistent with the phenomenology next.

  The $q^2$ dependence of $g_A(q^2)$ in the space-like $q^2$ region
is of a dipole form \cite{gA-exp}, namely,
\begin{eqnarray}
      g_A(q^2) &=& { {g_A(0)}\over {
   \left [1 - {q^2 / M_A^2} \right ]^2}},
\label{Exp-gA}
\end{eqnarray}
with $M_A \approx 1$ GeV. We shall use $M_A=1.0$ GeV in the following.

   The $q^2$ dependence of $g_{\pi NN}(q^2)$ is known less well than
that of $g_A(q^2)$. A monopole form for it, which can be parameterized
as
\begin{eqnarray}
      g_{\pi NN}(q^2) &=& g_{\pi NN}
        {{\Lambda_{\pi NN}^2 - m_\pi^2}
        \over
        {\Lambda_{\pi NN}^2 - q^2}}
\label{Exp-gpiNN}
\end{eqnarray}
is agreed upon in the literature; the value for $\Lambda_{\pi NN}$
varies. It is found to be greater than 1.2 GeV in nucleon-nucleon
(NN) scattering and deuteron property studies.
It is in contradiction to the expectations of many chiral nucleon
models for the nucleon. Lattice QCD evaluation also indicates
a smaller one, namely, $\Lambda_{\pi NN} \sim 800$ MeV.
On the
phenomenological side, Goldberger-Treiman discrepancy study \cite{GTDisp},
$pp\pi^0$ v.s. $pn\pi^+$ coupling constant difference \cite{pppi0},
high energy $pp$ scattering \cite{ppscat} and charge exchange
reaction \cite{q-exch,micCal},
etc, support a value of $\Lambda_{\pi NN}$ close to
$800$ MeV. By introducing a second ``pion'' $\pi'$ with mass 1.3
GeV, $\Lambda_{\pi NN}$ can be chosen to be around $800$ MeV
without spoiling the fit to the NN scattering phase shifts and
deuteron properties \cite{NND}. This picture was later
justified by a microscopic computation in Ref. \cite{micCal}.

   The $q^2$ dependence of $f_\pi(q^2)$ is little known from direct
experimental observations. It can be extracted from the following time
ordered correlator
\begin{eqnarray}
    q_\mu f_\pi(q^2) \delta^{ab} &=&
           -{1\over Z_{\pi}} (q^2-m^2_\pi)
           \int d^4x e^{iq\cdot x} \null_{(+)}\mathopen{\langle 0\,|}
           T \phi^a_\pi(x)A^b_\mu(0)\mathclose{|\,0\rangle}_{(-)}.
\label{fpi-q2-1}
\end{eqnarray}

The pion is a composite particle in QCD. Eq. \ref{fpi-q2-1} can
nevertheless be constructed using the following procedure.
First, consider a three point
function with two quark fields (in the pionic channel) and one axial
vector current operator. It contains a pion pole at
$q^2=m_\pi^2$.  Second, solve, e.g., Bethe-Selpeter or whatever suitable
equation(s) to obtain the $q^2$ independent
vertex function (or the wave function) for a pion.
Third, project out the pion contribution to the three
point function at arbitrary $q^2$
by using certain orthogonality relation\footnote{The orthogonality relation
between vertex functions in the relativistic case may
need generalization. It is however expected to
exist and to be unique. So the projection procedure is not ambiguous.
In many analytic diagramatical calculations based on simple models,
the pion contribution term can
simplly be read out from the three point function.}
between the pion vertex functions and the other parts of the
quark--antiquark scattering amplitude (the four point function) in the
pionic channel. Therefore the correlator in the above equation can in
principle be
computed from the QCD Lagrangian. The result is also expected to be
unique. It is however hard in practice to obtain a
reasonable result since QCD has not been solved. We need to resort to more
controlable methods that connect to experimental data and is accurate
enough.

At low $q^2$, it is beneficial to express it
in terms of an once subtracted dispersion relation, namely,
\begin{eqnarray}
    f_{\pi}(q^2) = f_\pi(m_\pi^2) + {q^2 - m_\pi^2 \over \pi}
   \int_{s_{th}}^\infty ds' {Im f_\pi (s')\over (s'-q^2)
     (s'-m_\pi^2)}.
\label{fpi-q2-2}
\end{eqnarray}
This is because the resulting $f(q^2)$ does not depend on
detailed shape of $Im f_\pi(q^2)$ but only some low moments of it when
$q^2$ is sufficiently small.
The lightest physical state connects to the axial vector current
operator with the quantum number of pion is
the $\rho\pi$ two particle state, the value of $s_{th}$ is chosen to
be $(m_\rho + m_\pi)^2$, where $m_\rho = 770$ MeV. At $s'= 4 m_N^2$,
which correspond to the lowest invariant mass of a $\bar N N$ system,
another branch cut for $f_\pi(q^2)$ develops.
We shall include $\rho\pi$ state only since $\bar N N$
state contributions to the above equation  is small when $q^2$ is small.
Our next step involves the specification or computation of $Im
f_\pi(s)$ by exploring the fact that only the $\rho\pi$ state which couples
to the pion contributes to $Im f_\pi(s)$ in the momentum transfer
region of interest to this paper.
An one loop computation of $Im
f_\pi(s)$ is known to be insufficient to account for experimental data
in other studies \cite{NND,micCal}, the $\rho \pi$ correlation,
which forms a resonance near $1.3$ GeV, is required. We therefore
propose the following form for $Im f_\pi(s)$,
\begin{eqnarray}
    Im f_\pi(q^2) &=& {3\over 4} {m_N\over g_{\pi NN}}
    {g_{\rho\pi\pi}^2\over 4 \pi}\bar\rho_{\pi,\rho\pi}(q^2)
\label{Imfpi-q2-1}
\end{eqnarray}
with the reduced density of state
\begin{eqnarray}
    \bar\rho_{\pi,\rho\pi}(q^2) &=& \bar\rho_0(q^2) \left ( 1 + {\lambda I_B
      \over (q^2-s_B)^2 + \bar\rho^2_0(q^2) I_B^2}
     \right )
\label{state-dens-1}
\end{eqnarray}
and
\begin{eqnarray}
   \bar \rho_0(q^2) &=& \sqrt{1-{(m_\rho+m_\pi)^2\over q^2}}
    \sqrt{1 - {(m_\rho-m_\pi)^2\over q2}}
   \left (1 -{m_\rho^2-m_\pi^2\over 3 q^2}   \right )\nonumber\\
    && \hspace{0.5in}
   \theta[q^2-(m_\rho+m_\pi)^2],
\label{state-dens-2}
\end{eqnarray}
where $\theta(x)$ is the step function with a value of unity for
positive x,
$s_R$ is  chosen to be $1.69$ GeV$^2$, $\lambda$ characterizes
the strength of the $\rho \pi$ resonance in $Im f_\pi(q^2)$ and
$\bar\rho_0(s_R) I_B$ characterizes the width of the resonance.

The above form is chosen so that when $\lambda=0$, $Im f_\pi(q^2)$ is
the one loop result in the Feynman-t' Hooft gauge (for the $\rho$
propagator). The $\rho\pi\pi$ interaction piece of Lagrangian density
used for evaluation of $Imf_\pi(q^2)$ is
\begin{eqnarray}
      {\cal L}_{\rho\pi\pi} &=& g_{\rho\pi\pi}
          \epsilon^{abc} \pi^a \partial^\mu \pi^b \rho_\mu^c.
\label{Lag-rhopipi}
\end{eqnarray}
The corresponding piece of the
axial vector current operator, which can be obtained
from the Noether theorem, can be written as
\begin{eqnarray}
    A^a_\mu &=& g_{\rho\pi\pi} {m_N\over g_{\pi NN}}
    \epsilon^{abc}\pi^b\rho_\mu^c,
\end{eqnarray}
where use has been made of the linear $\sigma$ model
relation $m_N = g_{\pi NN}\sigma$.

The value for $\lambda$, $\Lambda_{\pi NN}$ and $I_B$ is
adjusted so that the minimum value of the following function
\begin{eqnarray}
  f(\lambda,\Lambda_{\pi NN},I_B) &=& {1\over N}\sum_{k=0}^N
  {\left [m_N g_A(q^2_k)-g_{\pi NN}(q^2_k)f_\pi(q^2_k)   \right ]^2
   \over m_N^2 g^2_A(q^2_k) } e^{3 q^2_k},
\label{f-fit}
\end{eqnarray}
with $q^2_k = q^2_{min} + k {q^2_{max}-q^2_{min}/ N}$,
is achieved. The value of $N$ is chosen to be 100. $q_{min}^2 =
-0.6$ GeV and $q^2_{max} = 0.2$ GeV. The factor $e^{3q^2}$ is
used to put more weight on small $|q^2|$ region where the fit tends to
be poor.

\begin{figure}[h]
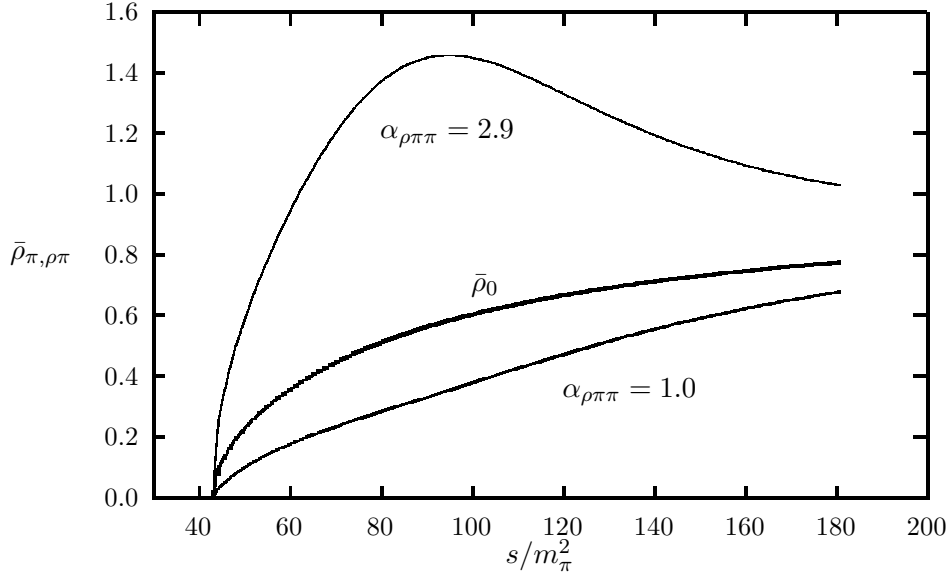

\begin{center}
\setlength{\unitlength}{0.240900pt}
\ifx\plotpoint\undefined\newsavebox{\plotpoint}\fi
\sbox{\plotpoint}{\rule[-0.500pt]{1.000pt}{1.000pt}}%

\end{center}
\caption{\label{Fig:PCAC} The $\rho\pi$ reduced density of states in
          the pionic channel
          as a function of $q^2$ obtained from the fitting procedure
          for two values of the $\rho\pi\pi$ coupling constant.
          For comparison, the one loop ($\lambda=0$)
          reduced density of state is
          drawn with a dotted line. Here $\alpha_{\rho\pi\pi}\equiv g^2_{\rho
           \pi\pi}/4\pi$.}
\end{figure}

In all cases studied and presented in table \ref{Table2},
\begin{table}[h]
\caption{The results of fitting, where $g_A=1.26$, $g_{\pi NN}= 13.4$,
          $f_\pi=93.2$ MeV,
         $q_{min}^2=-0.6$ $\mbox{GeV}^2$ and $q_{max}^2=0.2$ $\mbox{GeV}^2$.
         The unit for all but $\lambda$ and $g_{\rho\pi\pi}$ is GeV.
         $\lambda$ and $g_{\rho\pi\pi}$ are dimensionless.
         The quantities with a star on top is chosen by physical
         considerations.\label{Table2}}
\begin{center}
\begin{tabular}{|c|ccccccc|}
\hline
$g^2_{\rho\pi\pi}/4\pi$ & $\Lambda_{\pi NN}$ & $\lambda$ & $\sqrt{I_B}$ &
             $\sqrt{s_{R}^*}$ & $M_A^*$ & $\sqrt{s_{th}^*}$ &\\
\hline
1.0 & 0.94 &1.12 & 1.12 & 1.30 & 1.00 & 0.91 & \\
1.5 & 0.93 & 0.41 & 1.26 & 1.30 & 1.00 & 0.91 & \\
2.0 & 0.91 & -0.0023 & 1.34 & 1.30 & 1.00 & 0.91 & \\
2.5 & 0.89 & -0.25 & 1.27 & 1.30 & 1.00 & 0.91 & \\
2.9 & 0.88 & -0.39 & 1.30 & 1.30 & 1.00 & 0.91 & \\
3.5 & 0.86 & -0.56 & 1.35 & 1.30 & 1.00 & 0.91 & \\
\hline
\end{tabular}
\end{center}
\end{table}
$\Lambda_{\pi NN} < 0.95$ GeV. If a
value $g_{\rho \pi\pi}^2/4\pi = 1.0$ is taken, the qualitative shape
of $Imf_\pi(q^2)$ in Fig. \ref{Fig:PCAC} obtained by a minimization of Eq.
\ref{f-fit} is similar to the $Im\Gamma(q^2)$ of Ref. \cite{micCal}.
 Quantitatively, it has a broader width. The
phenomenological value for $g_{\rho\pi\pi}$ can be deduced from
the $\rho\to \pi \pi $ decay process. It has a value satisfies
$g_{\rho\pi\pi}^2/4\pi \approx 2.9$. Using this value of
$g_{\rho\pi\pi}$, $Imf_\pi(q^2)$ is obtained by minimization.
The result is given in Fig. \ref{Fig:PCAC}.
It's drastically different in shape from that of $Im\Gamma(q^2)$ in
Ref. \cite{micCal}. In fact, instead of increasing the density of
states relative to the one loop result, the resonance
contribution decreases the density of states in order to satisfy
Eq. \ref{Relation2}. The reduction of density of state indicates either the
solution is unphysical (for a normal resonance) or there is a
competing resonance in another channel that couples to the pionic
channel we are dealing with. But what can the ``other resonance''
channel be? There is no known resonance there. There seems to be
an inconsistency.

Of course this conclusion has to be checked by other
computations of $Imf_\pi(q^2)$ in different
model Lagrangians or in more fundamental ones like the lattice QCD calculation.
It is a worthy topic to be examined in the future.

\section{Extended PCAC relation for a nucleon}

   Two relationships between experimentally accessible nucleon form
factors $g_A(q^2)$, $g_P(q^2)$ and $g_{\pi NN}(q^2)$ and the less well
known $f_\pi(q^2)$ off the pion mass shell are established based on the facts
that 1) the QCD Lagrangian has an approximate $SU(2)_L\times SU(2)_R$
symmetry, which is explicitly broken down only by a small current mass
term (see Eq. \ref{QCD-Lag}) 2) this chiral symmetry is spontaneously
broken down to an flavor $SU(2)_V$ symmetry with pion, which is
lighter than it should be as an ordinary hadron, as the Goldstone
boson of the symmetry breaking 3) the pion, being lighter than other
normal hadronic particles, should dominate the low momentum transfer
reactions in certain channel, which is expressed as the PCAC relation
given by Eq. \ref{PCAC-QCD}.

   The first relation Eq. \ref{Relation1} passed the experimental test
using the available data. The second one seems to be in trouble when
it is confronted with our empirical knowledge. In general, if the
deviation is finally established, the remedy
for it can be either at a fundamental level, which means to modify the
contemporary field theoretical framework or QCD, or at the
structural level in the sense of revising our notion of what the
structure of the physical system under investigation is.
We adopt the second alternative in the following since even if the
unlikely possibility that the
inconsistency discussed in the above section is finally elliminated
after more detailed study, it is still an intellectual challenge how
the PCAC relation for a nucleon can be extended within the current
theoretical frame work. After all, it is hard to imagine the portion
of the vacuum state inside a nucleon remains unchanged under the
influence of the baryon density of the valence quarks that are
compacted into a region of order 1 fm.

The basic idea of the current approach to extend the PCAC relation
consists of a modification of the correspondance Eq.
\ref{PCAC-QCD} when its matrix elements are taken between nucleon
states  by assuming that
besides the pion, there is a different set of soft modes inside a
nucleon originating from
the spontaneous breaking of the chiral $SU(2)_L\times SU(2)_R$
symmetry down to the same flavor $SU(2)_V$ symmetry. Such a
possibility is theoretically investigated in Refs. \cite{YingL,YingA}
and are presented in the first part.
If there are such a set of these soft modes inside a nucleon, then the
left hand side (l.h.s.) of Eq. \ref{PCAC-QCD} in the low momentum
transfer regions is saturated not only by
the pions, but also these soft modes (Goldstone diquark excitations),
which exist if a localized $\beta$-phase is assumed.

These color carrying soft modes are confined inside the nucleon and
therefore can not be directly observed like the pions but can has an
effect by have non-pole contributions to Eq. \ref{Relation2}. The
result \cite{Epcac} is
\begin{eqnarray}
 m g_A(q^2) &=&
g_{\pi NN}(q^2) f_\pi(q^2) + (q^2 - m_\pi^2) \tilde
\eta (q^2), \label{PCAC5}
\end{eqnarray}
where $\tilde\eta(q^2)$ is a function related to the product of the
strength of the Goldstone diquark excitations and their
propagator, which has no pole at low energies (see Ref. \cite{GDH} for
a more careful discussion on this point) due to the fact that
they are confined inside the nucleon. It is worth mentioning that the
additional term has no effects on the pion mass shell $q^2-m_\pi^2$
due to the $q^2-m_\pi^2$ factor.

With this term added, the potential inconsistence between theory and
data discussed in the previous section can in principle be resolved.
It is clearly interesting to further investigate the existence or the
extend of the inconsistency discussed using various other models and
computation procedures.

\chapter{Gerasimov--Drell--Hearn Sum Rule for a Nucleon}

  The Gerasimov--Drell--Hearn (GDH) sum rule \cite{GMV,DH} is a
relation between the difference of two polarized total photon--
spin-1/2 Dirac
particle cross sections and the corresponding
particle's anomalous magnetic moment. It can be expressed as
\begin{eqnarray}
\int_{0}^\infty {\sigma_{3/2}(\nu) - \sigma_{1/2}(\nu)\over\nu} d\nu &=&
 {2\pi^2\alpha_{em}\over m^2}\kappa^2 ,
\label{GDH}
\end{eqnarray}
where $\alpha_{em}=1/137$ is the fine structure constant,
$\sigma_{3/2}$ is the total cross section when the photon
helicity is in the same direction as the target  particle's
spin polarization, $\sigma_{1/2}$ is the one when the photon helicity
antiparallels to the  particle's spin polarization and
$\kappa$ is the anomalous magnetic moment of the particle.

 It can be derived if the corresponding forward photon--particle
scattering amplitude decreases fast enough at high energies
so that an unsubtracted dispersion relation for the amplitude can be written
down. For the case of photon--nucleon scattering, the large energy
behavior can be inferred from phenomenology, which tells us that the
GDH sum rule ought to be satisfied. The current confrontation of the GDH sum
rule
for a nucleon to known experimental data from pion
photo-production result in disagreement \cite{WA,SWK}.
What causes the discrepancy is not yet clear.

Maybe the experimental information used to saturate the l.h.s.
of Eq. \ref{GDH} is not enough. It may also be that there is a true
need of an extension of the GDH sum rule on the theoretical part. This
study focus its attention to a theoretical study of the possibility of
an extension of the GDH sum rule consistent with Lorentz covariance and
local electromagnetic (EM) gauge invariance using the
field theoretical methods. We shall review briefly various issues concerning
the
modification of the GDH sum rule proposed. Details are given in
Refs. \cite{GDH,YingG}.

\section{Photon--nucleon compton scattering}

    The photon--nucleon forward Compton scattering amplitude is
related to the following covariant one photon irreducible
correlator between EM current operator $J^\mu(x)$
\begin{eqnarray}
     T^{\mu\nu}(p,q) &=& i\int d^4x e^{i q\cdot x}
                 \bra{pS} T^* J^\mu(x) J^\nu(0)\ket{pS},\label{Correlator1}
\end{eqnarray}
where $\ket{pS}$ is a nucleon state with 4-momentum $p^\mu$, polarization
$S^\mu$ and $T^*$ represents time ordering with proper Schwinger terms
added. Taking into account of the
conservation of parity in the EM interaction, the amplitude can be
parameterized by eight invariant amplitudes $F_{1\ldots 8}$
the following way
\begin{eqnarray}
     T^{\mu\nu}(p,q) &=& {1\over 2 m}
                      \bar U(pS)\left [F_1 g^{\mu\nu} - F_2 q^\mu q^\nu + F_3
                       p^\mu p^\nu - F_4 \left (p^\mu q^\nu + p^\nu q^\mu
                       \right )
                       + i F_5 \sigma^{\mu\nu}\right .  \nonumber \\ &&
                       + \left .
                            i F_6 \left (p^\mu\sigma^{\nu\alpha} q_\alpha -
                            p^\nu\sigma^{\mu\alpha} q_\alpha \right )
                       + i F_7 \left (q^\mu \sigma^{\nu\alpha} q_\alpha -
                            q^\nu\sigma^{\mu\alpha} q_\alpha \right )
                       + i F_8\epsilon^{\mu\nu\alpha\beta} q_\alpha
                       p_\beta {\rlap\slash q}\gamma^5
                          \right ] U(pS).\nonumber \\ &&
\label{Amplitude1}
\end{eqnarray}
The invariance amplitudes $F_{1\ldots 8}$ are functions of $q^2$ and
$\nu=p\cdot q/m$ with $m$ the mass of a nucleon.

Since the amplitude $T^{\mu\nu}$ satisfies the Ward identity
\begin{eqnarray}
     q_\mu T^{\mu\nu}(p,q) &=& 0 \label{T-ward}
\end{eqnarray}
due to the gauge invariance and commutativity of the EM charge density
operators at equal-time, it is usually written in a reduced form,
namely,
\begin{eqnarray}
    T^{\mu\nu}(p,q) &=&  S_1\left (-g^{\mu\nu} + {q^\mu q^\nu \over q^2} \right
)
                       + S_2 \left (p^\mu-{m\nu\over q^2} q^\mu\right)
                             \left (p^\nu-{m\nu\over q^2}
q^\nu\right)\nonumber\\
                    && - i A_1 \epsilon^{\mu\nu\alpha\beta}q_\alpha S_\beta
                       - i m \nu
                         A_2 \epsilon^{\mu\nu\alpha\beta}q_\alpha
                         \left (S_\beta - {S\cdot q \over m\nu} p_\beta \right
)
\label{Amplitude2}
\end{eqnarray}
with
\begin{eqnarray}
     S_1(q^2,\nu) &=& -F_1(q^2,\nu) = -q^2 F_2(q^2,\nu) - m \nu F_4(q^2,\nu)
     ,\label{S1-F}\\
     S_2(q^2,\nu) &=& F_3(q^2,\nu)
                  = {q^2\over m \nu} F_4(q^2,\nu),\label{S2-F}\\
     A_1(q^2,\nu) &=& m F_6(q^2,\nu) + \nu F_8(q^2,\nu),\label{A1-F}\\
     A_2(q^2,\nu) &=& {1 \over m} F_7(q^2,\nu) + F_8(q^2,\nu).\label{A2-F}
\end{eqnarray}

The forward cross section of the compton scattering is related to
the amplitude $M_{fi}$ given in to following
\begin{eqnarray}
     M_{fi} \sim {\epsilon'}^*_\mu \epsilon_\nu T^{\mu\nu}
\end{eqnarray}
where $\epsilon'$ and $\epsilon$ are the final and initial
photon polarizations.
In the nucleon rest frame, the polarization dependent part of the
forward cross section depends only on $A_1$ when the nucleon
polarization direction $S^\mu = (0,\vec{S})$ is along
direction of the photon propagation.

The invariant amplitude $A_1$ enjoys a sum rule given by Eq.
\ref{GDH}. Its derivation involves three issues discussed in the
following 1) is the use of the infinite momentum frame legitimate?
2) can the high energy
($\nu$) behavior of $A_1(q^2,\nu)$ be estimated using the Regge
asymptotics? 3) what's the role of gauge invariance in determining the large
$\nu$ behavior of $A_1(q^2,\nu)$?

\section{Sum rules and infinite momentum frame}

    The need of using an infinite momentum frame in deriving a fixed
$q^2$ sum rule like the GDH sum is discussed and emphasised in Ref.
\cite{AdlerBook}. It's validity in the derivation is generally
assumed. It is also assumed in this work.

\section{Regge behaviors and physical states}

    The large energy physical hadronic scattering amplitudes
follow certain power law (in energy) according to Regge asymptotics
\cite{ReggePh}, namely
\begin{eqnarray}
     \lim_{\nu\to\infty} A \sim \nu^{\alpha}
\end{eqnarray}
where $\nu$ is the energy variable and $\alpha$ is the leading Regge
trajectory of the corresponding channel. The highest trajectory is
that of the Permeron with an $\alpha(0)=1$ \cite{Permeron}. Since the
Permeron has a spin one (in the forward direction), it can not be
exchanged in some channels of reaction due to helicity conservation;
this allows the corresponding
amplitude to change slower than $\nu$. If the change is slow enough,
it allows certain superconvergence relation to be derived for it in
the corresponding channel for physical hadronic scattering amplitude
\cite{AdlerBook}.

The problems of using Regge asymptotic behavior in deriving sum rules
for matrix elements of a current (between physical hadronic states)
is known from current algebra studies \cite{BGLL,Singh}, in which it
was found that a hypothetical case of isovector photon--nucleon scattering
there appears the need of a ``J=1 fixed pole'' in the amplitude for
the matrix elements of the time ordered current--current correlator.
This problem can be understood if one realizes
that the matrix elements of the time ordered current--current
correlator does not correspond to a 4--points physical hadronic
scattering amplitude
directly. The phenomenological Regge asymptotics may fail for these
amplitudes. This situation opens the door for a modification of those
sum rule that are derived under the assumption of Regge behavior for
the matrix elements of the time ordered current--current
correlator. For the GDH sum rule, this is suggested in Refs.
\cite{AG,FGFD}.

For the realistic photon--nucleon scattering, this kind of ``J=1 fixed
pole'' effects can not be straight forwardly introduced in such a way
that the results are consistent with current algebra. This is
because it causes a modification of the commutation relation between
the time component of
the EM current operator at equal-time, which tells us
that they commute with each
other. This requirement, together with gauge invariance, are expressed
simply by Eq. \ref{T-ward}. A modification
of the commutation relation between the charge density operators at
equal-time is suggested in Refs. \cite{KS,CLW}. It leads to a
modification of the Ward identity Eq. \ref{T-ward}, so that a mutual
consistency of the arguments can be maintained. Whether or not such
a way of modifying GDH sum rule actually is consistent with
phenomenology remains to be shown however.

\section{Gauge invariance constraints}

The scheme proposed in this study is different in that the GDH sum
rule is modified by introducing an effective ``J=1 fixed pole'' effect
without modifying the commutation relation between the charge density
operators at equal-time. This is achieved by assuming there is a
localized breaking down of the $U(1)_{em}$ gauge symmetry inside a
nucleon \cite{GDH}. The ``J=1 fixed pole'' effects are provided by the
would-be Goldstone boson of the symmetry breaking,
which does not belong to physical excitation
spectrum for a local gauge symmetry \cite{WBG}.

{}From the Ward identity Eq. \ref{T-ward}, there are three constraints
for $F_{1\ldots 8}$ in Eq. \ref{Amplitude1}, namely,
\begin{eqnarray}
       F_1(q^2,\nu) - q^2 F_2(q^2,\nu) - m \nu F_4(q^2,\nu)
                                        & = & 0, \label{G-inv1}\\
       m \nu F_3(q^2,\nu) - q^2 F_4(q^2,\nu) & = & 0, \label{G-inv2}\\
       F_5(q^2,\nu) - m\nu F_6(q^2,\nu) - q^2 F_7(q^2,\nu)
                                 & = & 0. \label{G-inv3}
\end{eqnarray}
Eq. \ref{G-inv3} is relevant to the possibility of modifying the GDH
sum rule.

The large $\nu$ behavior of $F_5(q^2,\nu)$ can be determined to be
\cite{GDH} $F_5\sim \nu^{-1}$. If the following assumption is made,
namely,
\begin{eqnarray}
      \lim_{q^2\to 0} q^2 F_7 = 0,
\label{Assump1}
\end{eqnarray}
then the large $\nu$ behavior of $F_6$ at $q^2=0$ is $\nu^{-2}$
following Eq. \ref{G-inv3}. Under the assumption given by Eq. \ref{Assump1},
the large $\nu$ behavior of $A_1$, which is relevant to the
GDH sum rule, is controlled by that of $F_8$.
The large $\nu$ behavior of $F_8$
is bounded by the Regge asymptotics because it is an gauge invariant
invariant amplitude by itself, which means that it connects to
physical scattering amplitude only (for a detailed discussion on this
point, see Refs. \cite{GDH,YingG}). Thus $F_8\sim
A_1\sim \nu^{\alpha-1}$ with $\alpha<1$ from helicity amplitude
analysis \cite{Regge}. Thus gauge invariance requirement expressed by
Eq. \ref{T-ward} together with Eq. \ref{Assump1} eliminates the
possibility of a modification of the GDH sum rule.

\section{The possibility of modifying GDH sum rule}

The above analysis show that if GDH sum rule is to be modified in a
way that respect the local EM gauge invariance and the
Regge asymptotic behavior for physical hadronic scattering amplitudes,
assumption
Eq. \ref{Assump1} has to be relinquished by letting $F_7$ to has a
pole like behavior\footnote{Ref. \cite{GDH} provides a more precise meaning for
this
statement.} in $q^2$ at large $\nu$. This implies that the $U(1)_{em}$
symmetry corresponding to EM is spontaneously broken down inside a
nucleon \cite{GDH}. More specifically, introducing an order
parameter $\rho_\infty$, defined as
\begin{eqnarray}
        \rho_\infty
         &=& -\lim_{q^2\to 0}\lim_{\nu \to \infty} {q^2 F_7\over \nu
           }\label{orderpara}
\end{eqnarray}
for the EM gauge symmetry inside a nucleon\footnote{The first limit
$\nu\to\infty$ is necessary since localized
spontaneous gauge symmetry breaking is discussed here. A finite
region in space only looks more and more like an infinite system when
smaller and smaller distances or higher and higher energies
are probed. }. If the $U(1)_{em}$ gauge symmetry is spontaneously
broken down, then there is a massless pole in $F_7$ in the
$\nu\to\infty$ limit so that
$\rho_\infty$ can be nonzero. With $\rho_\infty$,
the GDH sum rule is modified to
\begin{eqnarray}
 \int_{0}^\infty {\sigma_{3/2}(\nu) - \sigma_{1/2}(\nu)\over\nu} d\nu =
 {2\pi^2\alpha_{em}\over m^2}\left (\kappa^2 + 2 m^2\rho_{\infty}\right ).
\label{ModGDH}
\end{eqnarray}

   Numerically, the value for $\rho_\infty^p$ and $\rho_\infty^n$ and
their difference can be extracted from the results of the integration of
the total photon--nucleon cross sections given in Ref. \cite{SWK}. They
take the value $\rho_\infty^p = 2.9\times 10^{-2} fm^2$ and
$\rho_\infty^n= -2.5\times 10^{-2} fm^2$. As a
final remark for this section, it should be mentioned that
$\rho_\infty^{p,n}$ has a dimension of length squared.
It can be written as $\rho_\infty^{p,n}=1/\Lambda^2$. It is interesting
to note that $\Lambda\sim 1 GeV$, which correspond to the natural scale
of the spontaneous chiral symmetry breaking in strong interaction.

It shows that if the current discrepancy between cross section
$\sigma_{3/2}$ and $\sigma_{1/2}$ obtained from the the photo pion
production on a nucleon data and the GDH sum rule is genuine, then a
localized spontaneous breaking down of the $U(1)_{em}$ symmetry inside
a nucleon has to be introduced. As it is discussed in part I of this
report, such a symmetry breaking is in principle possible since the
$U(1)_{em}$ gauge symmetry is spontaneously broken down in the
$\beta$- and $\omega$- phases \cite{YingA,YingL}.

\chapter{Rapidity Correlation of $B\bar B$ in High Energy
             $e^+e^-$ Annihilation}

   The hadronization processes in the $e^+e^-$ annihilation can be
described reasonably well by a chain like picture in terms of string
(or flux tube) fragmentation. For the meson
production, quark and antiquark pair is created by a breaking of the
string that connects the parent quark--antiquark pairs.
At the breaking point, any nonvanishing conserved quantum number like
charge, baryon number, etc. are created in
conjugate pairs so that they cancel locally (in space-time) there.
The reason behind it is that the string is considered to
be neutral in these quantum numbers.
Monte Carlo simulation programs like the JETSET
\cite{jetset} and Herwig \cite{herwig}, which are based upon such a
picture describe the experimental data well.

   This picture of hadronization implies strong rapidity correlation
between pairs of hadrons that conjugate to each other in the multihadron
final state. Since, before the string fragmentation,
it is in a stage of uniform expansion, the neighboring hadrons
produced on the string by a fragmentation has rapidity close to each
other than others. It is difficult to test the consequences of this
picture using the mesonic component of the
final state since 1) the correlating
meson pairs are hard to identify and 2) the interaction
between the mesonic particles are stronger than that of the baryons in
the $e^+e^-$ annihilation. The interaction distorts the rapidity
information of the meson at the time when it is created. On the other hand,
larger than 50$\%$ of the baryons escape the hadronic clouds before
interaction effects grow; they therefore carry the rapidity
information at their production time \cite{Xie}. One of the best characters of
the
string fragmentation picture for hadronization in the $e^+e^-$
annihilation processes is the strong rapidity
difference correlation between a pair of baryon and antibaryon ($B\bar
B$).

The experimental examination of it was carried out by observing the
$\Lambda\bar\Lambda$ rapidity difference correlation \cite{OPAL} in
$e^+ e^-$ annihilation at $\sqrt{s}=91$ GeV by OPAL collaboration.
A ``surprise'' was found that
the width of the rapidity difference correlation between $\Lambda\bar\Lambda$
pairs are wider than it is expected from the one predicted by the
above mentioned fragmentation model, which
consists of the creation of diquark--antidiquark pairs on the string
for the formation of baryons,
similar to the quark--antiquark pair creation for
mesons production. In order to describe the data, the so-called
popcorn mechanism
\cite{popcon} has to be introduced, and, in addition, the percentage
of the popcorn configuration has to be large in the fragmentation
processes \cite{Xie,OPAL}.

The popcorn mechnism implies a non-local production of
conjugate baryon number pairs on the string under the fragmentation
since, in between them, there has to be a finite space that contains
the meson. The conclusion that conjugate baryon numbers carried by diquark and
antidiquark on the string are not created at the same spacetime point
is hard to understand if the string that fragments into $B\bar B$
pairs are neutral in baryon number. According to the principle of
relativisty,
it violates the classical causality which requires local cancellation of
baryon number and antibaryon number at the time when they are created from a
baryon neutral source; the speed of their separation should be less
than the speed of light.

This difficulty can however be solved if one assumes that there exsits
$\beta$- or $\omega$- phase inside the string under the
fragmentation. The reason, as it is discussed in section \ref{sec:CP-vio}
of part I is that, spontaneous separation of baryon numbers are favored in
these phases (see, for example, Fig. \ref{Fig:Baryon}). Therefore if
one of the above mentioned phases is  present inside the
string before a fragmentation, the baryon density on the string can
be non-zero at a specific point; it is positive or negative alternatively
along the string.  Thus, by assuming the existence of
the $\beta$- or $\omega$- phase discussed in part I of the
report, we can gain a deeper understanding of the empirical
popcorn mechanism needed in phenomenology.

   Before closing this short chapter, let's reminded the reader that
this finding, if proved genuine in the future after more comprehensive studies,
could provide a domestic experimental basis\footnote{In the sense
that the physical processes underlying the mechnism can be created and
studied here, on earth.} for the
baryogensis mechanism in a matter--antimatter symmetric universe \cite{Omnes}
under the
standard cosmology (big-bang), which is based upon the spontaneous
separation of baryonic matter and antibaryonic matter at certain time
during the evolution of the univers \cite{YingL,YingB}, without
violating the observational constraints \cite{Steigman}. It is
certainly a worthy topic to be further studied.

\chapter{Summary and Outlook}

   The theoretically possible phases in an
interacting massless two flavor quark system are discussed in the
chapters of part I of this report by
introducing model Lagrangians. Four phases are found. Some of their
properties are discussed while others are mentioned with
references containing more details provided.
In part II, three different observations, which
is currently considered difficult to understand using the conventional
picture, are discussed in the light of the findings presented in part
I of this report. We find it likely that localized diquark condensed
phases exist inside a nucleon and hadronic excited states (within the flux
tube).

   In the heavy ion collisions, baryon number density and
the volume containing it can be much
larger than that of a nucleon. It is thus very probable that the
$\beta$- or $\omega$- phase can be produced despite the short
time duration of the collision. What signals their existence
existence and how to find them are questions that can be further
investigated.

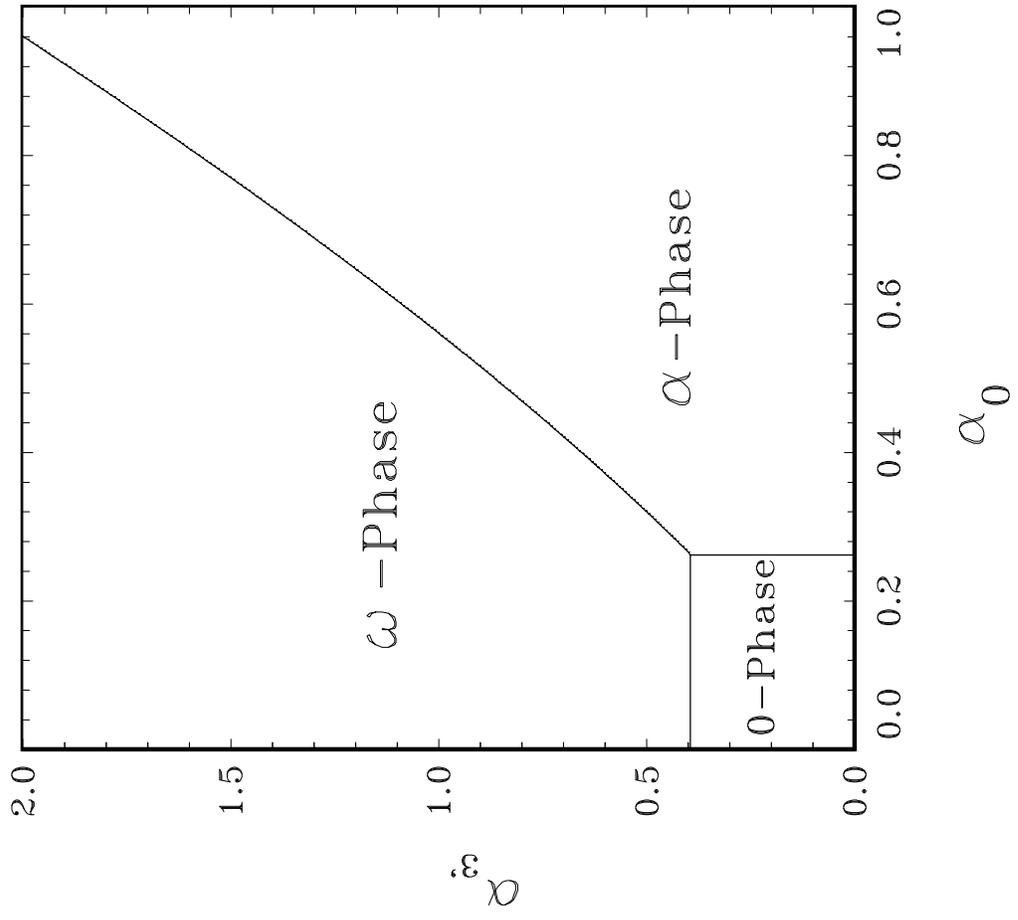
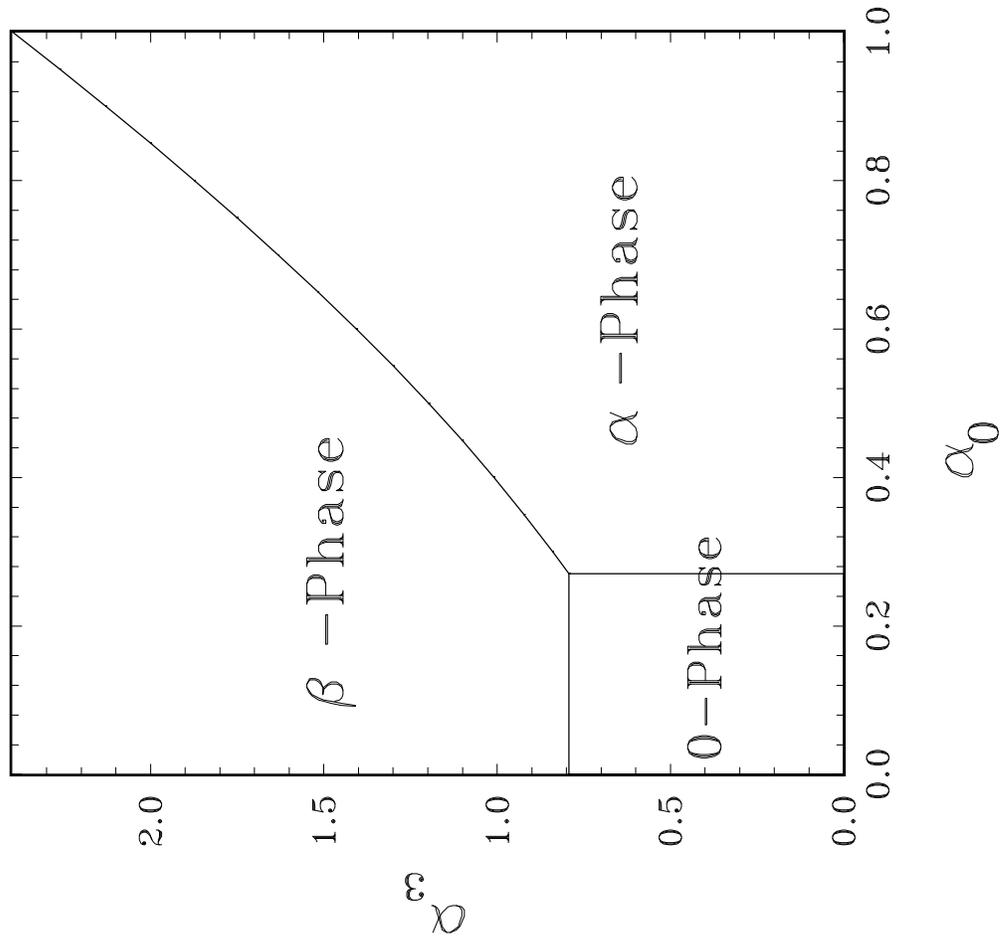
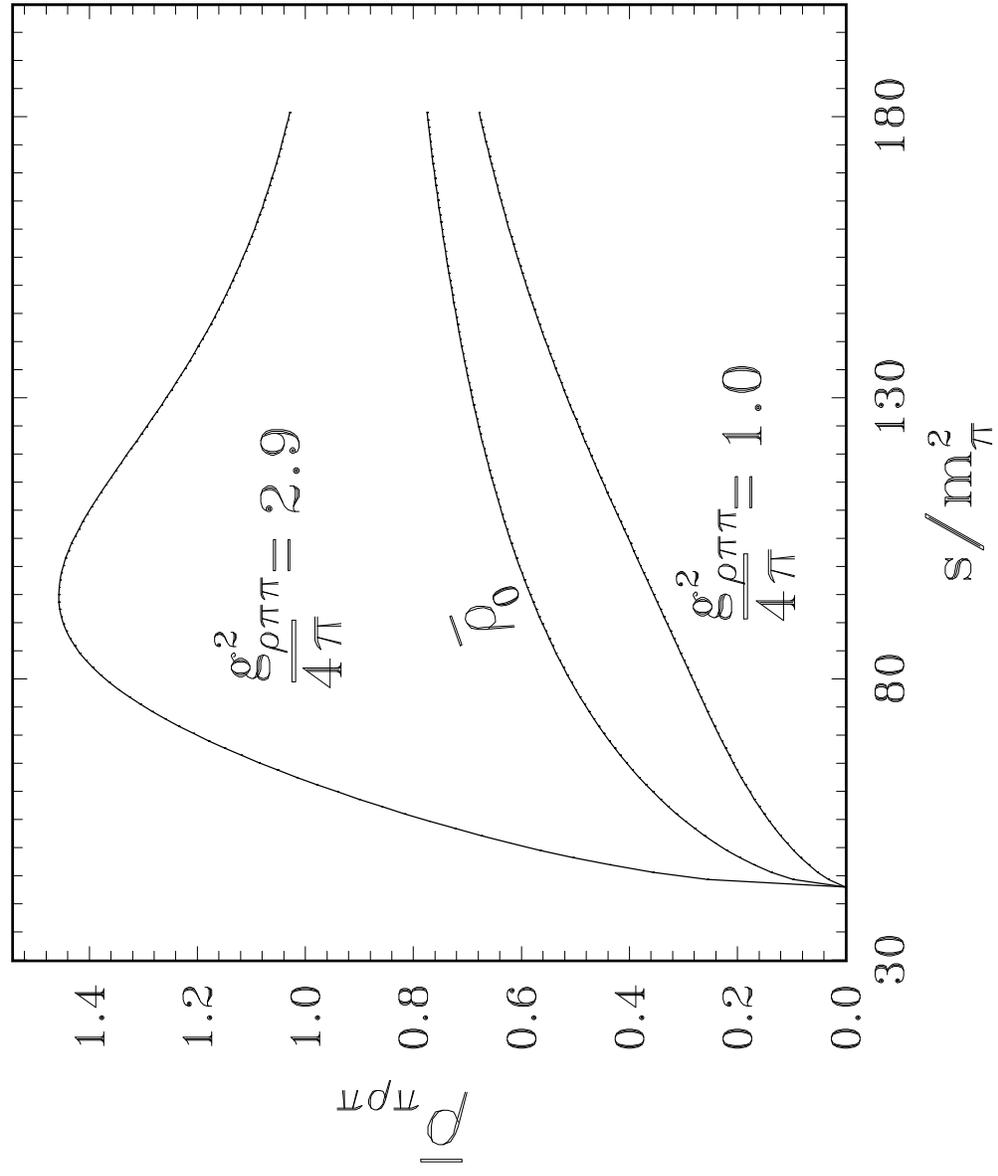
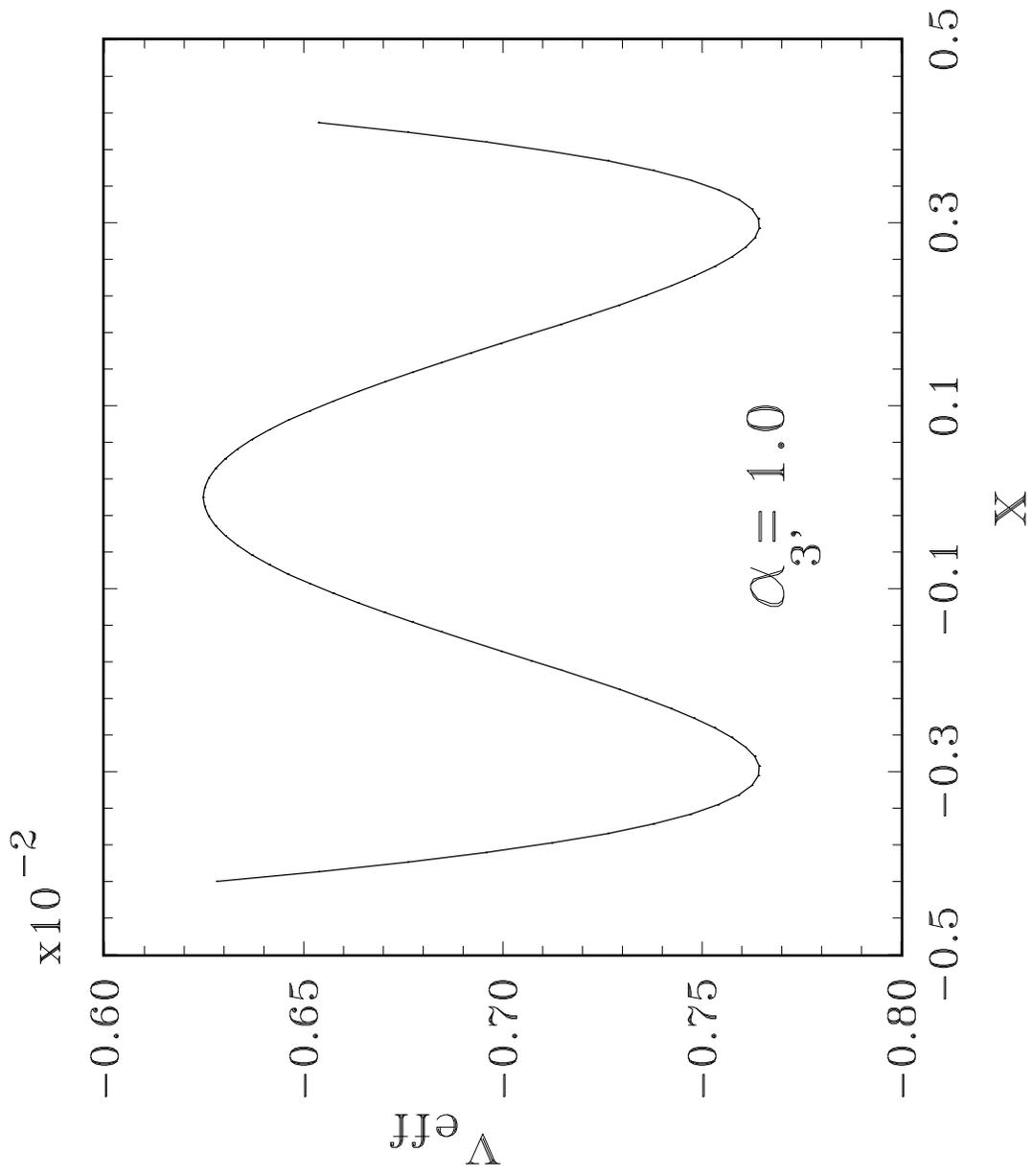
\end{document}